\documentclass[sn-basicespecial]{sn-jnl}%

\usepackage{cancel}

\theoremstyle{thmstyleone}%
\providecommand{\abs}[1]{\lvert#1\rvert}
\theoremstyle{thmstyletwo}%

\theoremstyle{thmstylethree}%

\bibpunct{(}{)}{;}{a}{ }{,}
\begin{document}

\title[Chaos, complexity, and intermittent turbulence in space plasmas]{Chaos, complexity, and intermittent turbulence in space plasmas}


\author{\fnm{A. C.-L. } \sur{ Chian}}
\author{\fnm{F. A.} \sur{ Borotto}}
\author{\fnm{T.} \sur{ Hada}}
\author{\\ \fnm{R. A.} \sur{ Miranda}}
\author{\fnm{P. R.} \sur{ Mu\~noz}}
\author{\fnm{E. L.} \sur{ Rempel}}



%
\abstract{Intermittent turbulence is key for understanding the stochastic nonlinear dynamics of space, astrophysical, and laboratory plasmas. We review the observation and theory of chaos and complexity in plasmas, and elucidate their links to intermittent plasma turbulence. First, we present evidence of magnetic reconnection and intermittent magnetic turbulence in coronal mass ejections in the solar corona and solar wind via remote and \textit{in situ} observations. The signatures of turbulent magnetic reconnection, i.e., bifurcated current sheet, reconnecting jet, parallel/anti-parallel Alfv\'en waves, and spiky dynamical pressure pulse, as well as fully-developed Kolmogorov intermittent turbulence, are detected at the leading edge of an interplanetary coronal mass ejection and the interface region of two interplanetary magnetic flux ropes. Methods for quantifying the degree of coherence, amplitude-phase synchronization, and multifractality of nonlinear multiscale fluctuations are discussed. The stochastic chaotic nature of Alfv\'en intermittent structures driven by magnetic reconnection is determined by a complexity-entropy analysis. Next, we discuss the theory of chaos, intermittency, and complexity for nonlinear Alfv\'en waves, and parametric decay and modulational wave-wave interactions, in the absence/presence of noise. The transition from order to chaos is studied using the bifurcation diagram. Two types of plasma chaos are considered: type-I Pomeau-Manneville intermittency and crisis-induced intermittency. The role of transient structures known as chaotic saddles in deterministic and stochastic chaos in plasmas is investigated. Alfv\'en complexity associated with noise-induced intermittency, in the presence of multistability, is studied. Finally, we discuss the relation of chaos, complexity, and intermittent turbulence in space plasmas to similar phenomena observed in astrophysical and laboratory plasmas, e.g., coronal mass ejections and flares in the stellar-exoplanetary environment and Galactic Center, as well as chaos, magnetic reconnection, and intermittent turbulence in laser-plasma and nuclear fusion experiments.}

\maketitle
\section{Introduction }  
Solar-terrestrial environment is a complex, electrodynamically coupled system dominated by stochastic nonlinear dynamical interactions  (Clemmow and Dougherty \citeyear{clemmow1969electrodynamics}; Chian et al. \citeyear{chian2006chaotic}; Kamide and Chian \citeyear{kamide2007handbook}; Miranda et al. \citeyear{miranda2021complexity}). The complex dynamics of solar-terrestrial plasmas, e.g., coronal mass ejections, solar flares, and geomagnetic storms, are an indication that the space plasma system is in a state far from equilibrium whereby instabilities, nonlinear waves, and turbulence play fundamental roles in the system dynamics. One of the ubiquitous features of complex space plasmas is the occurrence of intermittency and stochastic chaos in solar dynamo, solar corona, solar wind, and planetary magnetosphere-ionosphere-atmosphere. Intermittency is characterized by abrupt changes of the physical variables in time and/or space, e.g., the temporal variability of solar cycles with alternating periods of quiescent low-level fluctuations (solar minima) and bursting high-level fluctuations (solar maxima) interspersed with random occurrence of grand minima; it displays multiscale behaviors with power-law spectrum in frequency and wave number, as well as non-Gaussian statistics in the probability distribution function of fluctuations. Chaos is characterized by aperiodicity in time and/or irregularity in space; chaotic dynamical systems are sensitive to small changes in the initial conditions and system parameters, including noise. The aim of this paper is to present an overview of observation and theory of intermittency and stochastic chaos in space plasma turbulence.\\

Magnetic reconnection is a major physical mechanism that destabilizes solar coronal magnetic flux rope structures (Keppens et al. \citeyear{keppens2019ideal}). Since the solar atmosphere is permeated by myriads of magnetic coronal loops, the interaction of multiple magnetic flux ropes can lead to thin current sheets that are susceptible to magnetic reconnection, resulting in coronal mass ejections, solar flares, and electromagnetic emissions such as coherent radio bursts driven by particle acceleration (Melrose \citeyear{melrose2017coherent}). Multiscale nonlinear dynamics of solar prominences, associated with the magnetic Rayleigh-Taylor instability, can be responsible for plumes and prominence eruption (Hillier \citeyear{hillier2018magnetic}). The near-Earth space is a unique laboratory for investigating the role of small-scale coherent structures in energy dissipation processes in plasma turbulence at magnetohydrodynamics (MHD) and sub-ion (kinetic) scales in magnetized plasmas, thanks to the availability of high-quality spacecraft data (Sahraoui, Hadid and Huang 2020). Solar wind particles reflected from the Earth's bow shock produce waves and instabilities that can heat plasma and accelerate particles, resulting in complex upstream structures such as Short Large Amplitude Magnetic Structures (SLAMS), hot flow anomalies, and density holes (Parks et al. 2017). Solitary structures can evolve from nonlinear ion-acoustic waves generated by field-aligned shear ion flow and parallel current in auroral ionospheric plasmas (Salem and Ali Shan \citeyear{saleem2020theoretical}). Equatorial plasma depletions in the ionosphere (Farley et al. \citeyear{farley1970equatorial}; Booker \citeyear{booker1956turbulence} ) that have significant impact on space weather, causing rapid fluctuations in radio signals used in telecommunications, show complex characteristics of intermittent turbulence, e.g., non-Gaussianity, intermittency, multifractality, and amplitude-phase synchronization in multiscale interactions (Chian et al. \citeyear{chian2018multi}).\\

Nonlinear wave-wave interactions have been observed in laser-plasma experiments; above certain threshold of laser power, nonlinear processes such as parametric decay, stimulated scattering and filamentation are excited (Kaw \citeyear{kaw2017nonlinear}). Super-nonlinear periodic waves and solitons in the shear Alfv\'en and ion-acoustic modes have been observed in multi-species plasma experiments in the laboratory; these large-amplitude and long-period waves correspond to the outermost phase trajectories enveloping the separatrix whose total energy is above a certain Sagdeev pseudo-potential barrier height and the amplitude cannot be smaller than that of the separatrix (Dubinov and Kolotkov \citeyear{dubinov2018above}).  Various types of vortices in magnetized partially ionized plasmas have been observed in laboratory experiments, including plasma hole (vortex with a density hole), spiral vortex, tripolar vortex, and counter $\mathbf{E} \times \mathbf{B} $ vortex. Theoretical, numerical simulation, and observational evidence of nonlinear processes of kinetic Alfv\'en waves formulated by nonlinear gyrokinetic theory have been obtained (Chen, Zonca and Lin \citeyear{chen2021physics}), e.g., three-wave parametric decay instabilities, modulational instabilities associated with the spontaneous generation of convective cells, and the quasi-linear phase-space transport induced by kinetic Alfv\'en waves. Interaction between a magnetic island and turbulence has been studied in various plasma experiment devices. Plasma turbulence becomes strongly inhomogeneous around a magnetic island due to the combined effect of the pressure gradient and flow shear modifications by the island (Choi \citeyear{choi2021interaction}); complex turbulence phenomena can affect the island stability in fusion plasmas as well as the underlying magnetic reconnection process, e.g., turbulence spreading, nonlinear mode coupling, and turbulence-driven flow; thus turbulence can either suppress or facilitate the growth of magnetic islands. \\

The outline of this paper is as follows. In Section 2, we discuss the basic concepts of chaos, complexity, and intermittent turbulence. In Section 3, we discuss magnetic reconnection and intermittent turbulence in a coronal mass ejection associated with an erupting solar flare observed in the solar corona remotely by radio and EUV images, and in interplanetary coronal mass ejections observed \textit{in situ} by multi-spacecraft in the solar wind at 1 AU and at the Earth's bow shock. In Section 4, we apply the chaos theory to explain crisis-induced intermittency in the absence of noise and noise-induced intermittency in the presence of noise. The crucial role of chaotic saddles in generating both types of intermittency is clarified. In Section 5, we discuss the relation between stochastic nonlinear dynamical phenomena observed in solar-terrestrial plasmas and similar phenomena observed in astrophysical and laboratory plasmas. 
\section{Basic concepts}

    \subsection{Chaos} 
    The first observation of transition from order to chaos in the solar wind was reported by Burlaga (\citeyear{burlaga1988period}), who identified the formation of ordered large structures from irregular small structures as well as the period-doubling of the period of the corotating interaction regions in the outer heliosphere. Chian, Borotto and Gonzalez (\citeyear{chian1998alfven}) demonstrated that Alfv\'en intermittency can be driven by chaos in the solar wind. Chian et al. (\citeyear{chian2006chaotic}) studied the chaotic nature of the solar-terrestrial environment. Hanslmeier (\citeyear{hanslmeier2020book}) discussed the chaotic behavior of solar cycles.\\
    
    Chaotic systems exhibit various types of intermittency (e.g., Ott \citeyear{ott1993chaos}; Chian \citeyear{chian2007complex}). The intermittent route to chaos was discovered by Manneville and Pomeau (\citeyear{manneville1979intermittency}) who showed that the type-I Pomeau-Maneville intermittency, related to episodic regime switching between periodic and chaotic behaviors, occurs via a local bifurcation termed saddle-node bifurcation. Another chaotic scenario that leads to intermittency occurs when the system undergoes a global bifurcation termed crisis, whereby a chaotic attractor in the state space suddenly changes in size (interior crisis), disappears (boundary crisis) or two or more chaotic attractors merge to form a large chaotic attractor (attractor merging crisis) (Borotto et al. \citeyear{borotto2004alfven}; Chian et al. \citeyear{chian2005attractor}). After an interior crisis, crisis-induced intermittency appears involving episodic regime switching between periods of weakly and strongly chaotic behaviors. Intermittency also takes place after an attractor merging crisis (Rempel et al. \citeyear{rempel2005intermittency}).\\
    
    Stable and unstable periodic orbits are the building blocks of dynamical systems, and are key to understand the origin of intermittency. A dissipative dynamical system consists of order and chaos; order is characterized by a maximum Lyapunov exponent smaller than or equal to zero (Wolf et al. \citeyear{wolf1985determining}) and governed by stable equilibrium points, stable periodic orbits (limit cycles) and quasiperioric attractors, whereas chaos is characterized by a positive maximum Lyapunov exponent and governed by a chaotic set composed by an infinity of unstable periodic orbits. Unstable periodic orbits are the skeleton of chaotic attractors and chaotic saddles. Chaotic saddles (Lai and T\'el \citeyear{lai2011transient}) are non-attracting chaotic sets responsible for the chaotic transient inside a periodic window and for chaos and intermittency in the chaotic regions outside a periodic window, e.g., after a saddle-node bifurcation (type-I Pomeau-Manneville intermittency) and after an interior crisis (crisis-induced intermittency). Chaotic saddles are given by the intersections of stable and unstable manifolds, which have been observed in solar supergranular turbulence (Chian et al. \citeyear{chian2020lagrangian}). In practice, the observation of space plasma turbulence contains an admixture of chaos and noise. Noise can have contributions from other plasma processes and/or instruments. For example in the presence of noise, Alfv\'en extrinsic intermittency can be driven (Rempel et al. \citeyear{rempel2006alfven}, \citeyear{rempel2008alfven}). Hence,  in order to give a proper interpretation of the complex behaviour of space plasma turbulence, the chaos theory needs to be modified to take into account the effects of noise to describe accurately the observation of stochastic chaotic fluctuations (Miranda et al. \citeyear{miranda2021complexity}).\\
    
    Traveling wave solution provides a convenient way to obtain an insight of nonlinear waves in plasmas (Chian and Clemmow \citeyear{chian1975nonlinear}). In this paper, we will adopt the low-dimensional deterministic and stochastic chaos approach to discuss the nonlinear dynamics of Alfv\'en waves and nonlinear wave-wave interactions by seeking traveling wave solutions which transform a nonlinear partial differential equation to a set of nonlinearly coupled ordinary differential equations (Hada et al. \citeyear{hada1990chaos}; Chian et al. \citeyear{chian2000chaotic}; Miranda et al. \citeyear{miranda2013universal}). This approach allows us to interpret the observation of stochastic chaotic fluctuations in intermittent turbulence in space plasmas in terms of crisis-induced intermittency (Chian, Borotto and Gonzalez \citeyear{chian1998alfven}) or noise-induced intermittency  (Rempel et al. \citeyear{rempel2008alfven}). The information gained from low-dimensional chaos is the basis for understanding high-dimensional chaotic phenomena described by extended spatiotemporal dynamical systems, e.g., laminar-turbulent transition and edge of chaos (He and Chian \citeyear{he2003off}; Rempel and Chian \citeyear{rempel2007origin}; Chian et al. \citeyear{chian2010amplitude}; Chian, Mu\~noz and Rempel \citeyear{chian2013edge}). 
    
    \subsection{Complexity}  
    
    Quantification of the complex dynamics of space plasma turbulence can be carried out by measuring the degree of simplicity-complexity (Rempel et al. \citeyear{rempel2008alfven}), coherence-incoherence (Pikovsky, Robsenblum and Kurths \citeyear{pikovsky2003synchronization}; He and Chian \citeyear{he2003off}; Chian et al.  \citeyear{chian2010amplitude}), order-randomness (Aschwanden et al. \citeyear{aschwanden2018order}), and complexity-entropy (Bandt and Pompe \citeyear{bandt2002permutation}; Rosso et al. \citeyear{rosso2007distinguishing}; Miranda et al. \citeyear{miranda2021complexity}).\\
    
    Multistability refers to the simultaneous presence of more than one attractor for a given value of the system control parameter, and it can be an obstacle for prediction, since the asymptotic state may depend crucially on the initial condition (Rempel et al. \citeyear{rempel2008alfven}). Hence, information about the coexisting attractors and their respective basins of attraction is crucial for understanding the dynamics of multistable systems. Multistability is also related to the characterization of complex systems which often involve the features of (Crutchfield and Young \citeyear{crutchfield1989inferring}; Poon and Grebogi \citeyear{poon1995controlling}; Badii and Politi \citeyear{badii1997complexity}): (1) many parts that are related in a complicated manner; (2) coexisting simple (ordered) and complex (disordered) behaviours, and (3) structures with different temporal and/or spatial scales (multiscale systems). Rempel et al. (\citeyear{rempel2008alfven}) showed that these three features can be found in a nonlinear model of Alfv\'en waves subject to external noise; Rempel et al. (\citeyear{rempel2006alfven}) studied the occurrence of extrinsic transients (stochastic chaotic saddles), attractor hopping in a multistable Alfv\'en system due to the effect of an additive noise, and Alfv\'en noise-induced intermittency.\\
    
    Self-organization is a characteristic of dissipative nonlinear processes governed by a global driving force and a local positive feedback mechanism, which generate regular geometric and/or temporal patterns, and increase the degree of order locally by decreasing entropy, in contrast to random processes (Aschwanden et al. \citeyear{aschwanden2018order}).  Hence, self-organizing systems create spontaneous order out of randomness during the evolution from an initially disordered system to an ordered quasi-stationary system, mostly by quasi-periodic dynamics  or harmonic resonances. Various types of global driving force can induce self-organization, e.g., mechanical forces of rotation such as accretion discs or differential rotation such as stellar dynamo, while the positive feedback mechanism is often an instability, e.g., the magnetoconvective Rayleigh-B\'enard instability or turbulent magnetic reconnection. Chang (\citeyear{chang1992low}) showed that the self-organized critical approach can describe spontaneous evolution of a complex system to critical states modelled by low-dimensional stochastic dynamical systems. Valdivia et al. (\citeyear{valdivia2003self}) and Pulkkinen et al. (\citeyear{pulkkinen2006role}) confirmed that this approach provides a better representation for describing the complexity of magnetospheric systems than the low-dimensional deterministic chaotic approach.\\
    
    Synchronization is a physical mechanism by which nonlinear multiscale fluctuations self-organize to exhibit varying degrees of coherence-incoherence in space and/or time (Pikovsky, Rosenblum, and Kurths,  \citeyear{pikovsky2003synchronization}). Hada et al. (\citeyear{hada2003phase}) developed a surrogate data technique and introduced a phase coherence index to quantify the degree of phase synchronization in plasma turbulence. He and Chian (\citeyear{he2003off}) identified a new phenomenon of phase synchronization in a type of on-off spatiotemporal intermittency in a well-developed drift-wave plasma turbulence; in ``on'' stages, the oscillators in different spatial scales adjust themselves to collective imperfect phase synchronization, inducing spiky soliton-like phase-coherent bursts in the wave energy. Chian et al. (\citeyear{chian2010amplitude}) demonstrated the duality of amplitude and phase synchronization related to spatiotemporal multiscale interactions in chaotic saddles at the onset of permanent spatiotemporal chaos in a nonlinear model of drift plasma waves using the Fourier-Lyapunov representation; the computed time-averaged Fourier power and phase spectral entropy showed that the laminar (bursty) state in the on-off spatiotemporal intermittency, after the laminar-turbulent transition via an interior crisis, corresponds to weak (strong) chaotic saddle with higher (lower) degree of amplitude-phase synchronization across spatial scales.\\
    
    Chaotic systems share some common properties with stochastic processes such as wide-band power spectrum, a delta-like autocorrelation function, and irregular behavior of the measured signal that make them difficult to be distinguished (Maggs and Morales \citeyear{maggs2013permutation}). Rosso et al. (\citeyear{rosso2007distinguishing}) introduced the Jensen-Shannon complexity-entropy plane to allow the distinction of chaos from noise. The vertical and horizontal axis in the complexity-entropy plane are suitable functionals of the pertinent probability distribution, namely, an appropriate statistical complexity measure and entropy of the system, respectively. The statistical characteristics of a time series can be determined by obtaining its statistical complexity and permutation entropy, which can be computed from a probability distribution introduced by Bandt and Pompe (\citeyear{bandt2002permutation}). This complexity-entropy technique has been applied to laboratory and space plasma turbulence (Maggs-Morales \citeyear{maggs2013permutation}, \citeyear{maggs2015chaotic}; Miranda et al. \citeyear{miranda2021complexity}).
    
    \subsection{Intermittent turbulence}   
    
Intermittent turbulence is associated with a non-Gaussian probability distribution function of fluctuations with fat-tails at small-scales which are the signature of extreme events, sharp spikes in the time series of fluctuations indicative of intermittent coherent structures at small-scales, and power-law power density spectrum with breaking manifesting turbulent cascade linked to nonlinear wave-wave interactions and dissipation (Frisch \citeyear{frisch1995turbulence}). Multifractality in intermittent turbulence can be characterized by a departure of the scaling exponent of the moments of fluctuations from the self-similarity assumption made by Kolmogorov in 1941 for an isotropic homogeneous turbulence. Burlaga (\citeyear{burlaga1991multifractal}) was the first to report the evidence of intermittent turbulence in solar wind by identifying the existence of multifractal structures in the velocity fluctuations associated with recurrent streams at 1 AU and near 6 AU. Solar wind turbulence consisting of Alfv\'enic fluctuations and convected magnetic coherent structures, such as magnetic flux ropes and current sheets, are the driver of geomagnetic (auroral) activities (D'Amicis, Telloni and Bruno \citeyear{d2020effect}). The extensive literature on interplanetary intermittent turbulence has been reviewed by Bruno and Carbone (\citeyear{bruno2013solar}), Matthaeus et al. (\citeyear{matthaeus2015intermittency}), and Oughton and Engelbrecht (\citeyear{oughton2021solar}). \\

Magnetic reconnection involving topological changes in magnetic fields is a fundamental process in intermittent magnetic turbulence in space and laboratory plasmas, that can lead to energy release in regions of magnetic field annihilation; in strong turbulence, magnetic field lines constantly reconnect everywhere and on all scales, thus making magnetic reconnection an intrinsic part of turbulent cascade (Lazarian et al. \citeyear{lazarian20203d}). Lazarian and Vishniac (\citeyear{lazarian1999reconnection}) showed that turbulence leads to fast magnetic reconnection, thus magnetic reconnection and turbulence are intrinsically connected. Wei et al. (\citeyear{wei2003magnetic}) reported observational evidence of magnetic reconnection in various solar wind structures such as at magnetic cloud boundary layers, heliospheric current sheet, and small-scale turbulent structures; the basic characteristics of magnetic reconnection in interplanetary plasmas include multiple X-line reconnection, vortex velocity structures, filament current systems, splitting, collapse of bulk plasma, and merging of magnetic islands. Magnetic reconnection exhausts, at thin current sheets with moderate to large changes in magnetic field orientation, were detected by Gosling et al. (\citeyear{gosling2005direct}) in the interior of an interplanetary coronal mass ejection (ICME) and at the interface between two ICMEs; the prime evidence is the acceleration of ion flow within magnetic field reversal region which was consistent with the Wal\'en relationship relating changes in flow velocity to density-weighted changes in the magnetic field vector; pairs of proton beams along the magnetic field were observed near the center of the accelerated flow event; the resulting reconnecting jets occurred within a Petschek-type reconnection exhaust region bounded by Alfv\'en waves. The discovery of reconnection exhausts in the solar wind introduces a new laboratory where magnetic reconnection can be investigated by \textit{in situ} measurements using widely separated multi-spacecraft, which enabled the observation of a magnetic reconnection X-line extending more than 390 Earth radii in the solar wind (Phan et al. \citeyear{phan2006magnetic}); the abrupt changes in the magnetic field $B_z$ at the two edges and a plateau in the $B_z$ profile in the middle of the current sheet indicate that the current sheet is bifurcated; the plasma density and temperature were sharply enhanced at the edges of the current sheet while the magnetic field strength was reduced. Chian and Mu\~noz (\citeyear{chian2011detection}) and Chian et al. (\citeyear{chian2016genesis}) obtained observational evidence of fully-developed Kolmogorov intermittent magnetic turbulence in the region of bifurcated current sheets associated with magnetic reconnection exhausts at the leading edge of an ICME and at the interface of two merging interplanetary magnetic flux ropes, respectively. Chian et al. (\citeyear{chian2016genesis}) showed that the condition for occurrence of magnetic reconnection derived by Swisdak et al. (\citeyear{swisdak2010vector}), relating the jump in the plasma parameter across the current layer and the shear angle between the reconnecting magnetic fields, can be applied to identify the most likely site of magnetic reconnection in a region of multiple magnetic flux ropes in solar wind. \\

    Magnetic flux ropes are bundles of helical, current-carrying, magnetic field lines writhing about each other and spiraling around a common axis. These coherent structures are a key element of heliospheric dynamics, contributing to the acceleration and transport of suprathermal particles in the expanding turbulent solar wind (Ruffolo et al. \citeyear{ruffolo2013squeezing}). Magnetic flux ropes and current sheets are commonly used as tracers of magnetic reconnection. Such structures can be reconstructed by the Grad-Shafranov method which models the magnetic field structure traversed by a spacecraft (Hu et al. \citeyear{hu2004multiple}) and are useful for studying CME-CME merger and interplanetary rope-rope magnetic reconnection (Hu et al. \citeyear{hu2004multiple}; Chian et al. \citeyear{chian2016genesis}). Khabarova et al. (\citeyear{khabarova2016small}) showed that magnetic reconnection due to small-scale magnetic island merging and contraction can provide an effective mechanism for particle acceleration in solar wind. Li (\citeyear{li2008identifying}) developed a method of detecting current sheets formed from intermittent energy dissipation in nonlinear interactions in the multifractal solar wind turbulence, by studying the integrated distribution function of the angle between two-point correlation of the magnetic field; it is plausible that these current sheets are the magnetic walls of adjacent magnetic flux ropes in solar wind. Chian and Mu\~noz (\citeyear{chian2011detection}) applied the method of Li (\citeyear{li2008identifying}) to detect a large number of small-scale current sheets at the shock-sheath region of an ICME. \\
    
Data from new space missions have significant impact in improving our concepts of space plasma complexity. \textit{Parker Solar Probe} showed that large-amplitude, Alfv\'enic magnetic-field reversals known as magnetic switchbacks (Bale et al. \citeyear{bale2021solar}) are prevalent in the inner heliospheric plasmas; these spiky fluctuations occur over a range of timescales and in patches separated by intervals of quiet, radial magnetic field, typical of intermittent turbulence. Switchbacks are localized within the extensions of plasma structures originating at the coronal base; these structures are accompanied by an increase in alpha particle abundance, Mach number, plasma $\beta$ and pressure, and by decreases in the magnetic field magnitude and electron temperature; these intervals are in pressure balance, implying stationary spatial structure, and the magnetic-field decreases are in agreement with overexpanded magnetic flux tubes. In particular, these structures are separated in longitude by supergranular scales, which suggests that switchbacks originate near the leading edge of the diverging magnetic field funnels associated with the network magnetic field at supergranular junctions, namely, the primary sources of solar wind. The above observations enabled Bale et al. (\citeyear{bale2021solar}) to propose that switchbacks are driven by interchange magnetic reconnection events (with the footpoints of reconnecting closed and open magnetic flux ropes rooted at supergranular junctions) just above the solar transition region and the spacecraft measurements represent the extended regions of a turbulent outflow of magnetic reconnection exhaust. Fargette et al. (\citeyear{fargette2021characteristic}) concluded that switchbacks are formed in the low corona and modulated by the solar surface convection patterns of supergranulation and granulation; the large scales detected for switchback patches are compatible with supergranulation scales and the smaller scales are compatible with granulation scales.  Eastwood et al. (\citeyear{eastwood2021solar}) observed an ion-scale magnetic flux rope confined to a bifurcated current sheet within a magnetic reconnection exhaust in the solar wind using \textit{Solar Orbiter} and \textit{Wind} data, thus demonstrating that reconnection signatures can be found separated by as much as $\sim 2000$ Earth radii, or 0.08 AU. Froment et al. (\citeyear{froment2021direct}) reported evidence of magnetic reconnection occurring at the boundaries of three switchbacks crossed by \textit{Parker Solar Probe} at a distance of 45 to 48 solar radii to the Sun during its first encounter. Fedorov et al. (\citeyear{fedorov2021switchback}) used {Parker Solar Probe} and {Solar Orbiter} data, when the two spacecraft were located around the same Carrington longitude and their latitudinal separation was very small,  to study switchbacks in the solar wind originating from the same coronal hole region; Solar Orbiter observed bent magnetic field lines that reconnect with each other, producing flux ropes, which suggests that the observed magnetic flux ropes might be the surviving and modified remnants of the switchbacks created near Sun observed by Parker Solar Probe. Telloni et al. (\citeyear{telloni2021evolution}) used the first radial alignment data of \textit{Solar Orbiter}-\textit{Parker Solar Probe} to investigate the radial evolution of solar wind turbulence in the inner heliosphere; two 1.5 h intervals of the magnetic field data were used to compute the power spectral density, flatness, and high-order moment scaling law; the results show that solar wind plasma evolves from a highly Alfv\'enic, less-developed turbulence state near the Sun, to fully-developed intermittent turbulence at 1 AU. 

\section{Observation}  

    \subsection{Coronal mass ejection and interplanetary coronal mass ejection} 
    \begin{figure}[ht]
        \centering
        \includegraphics[width=0.8\linewidth]{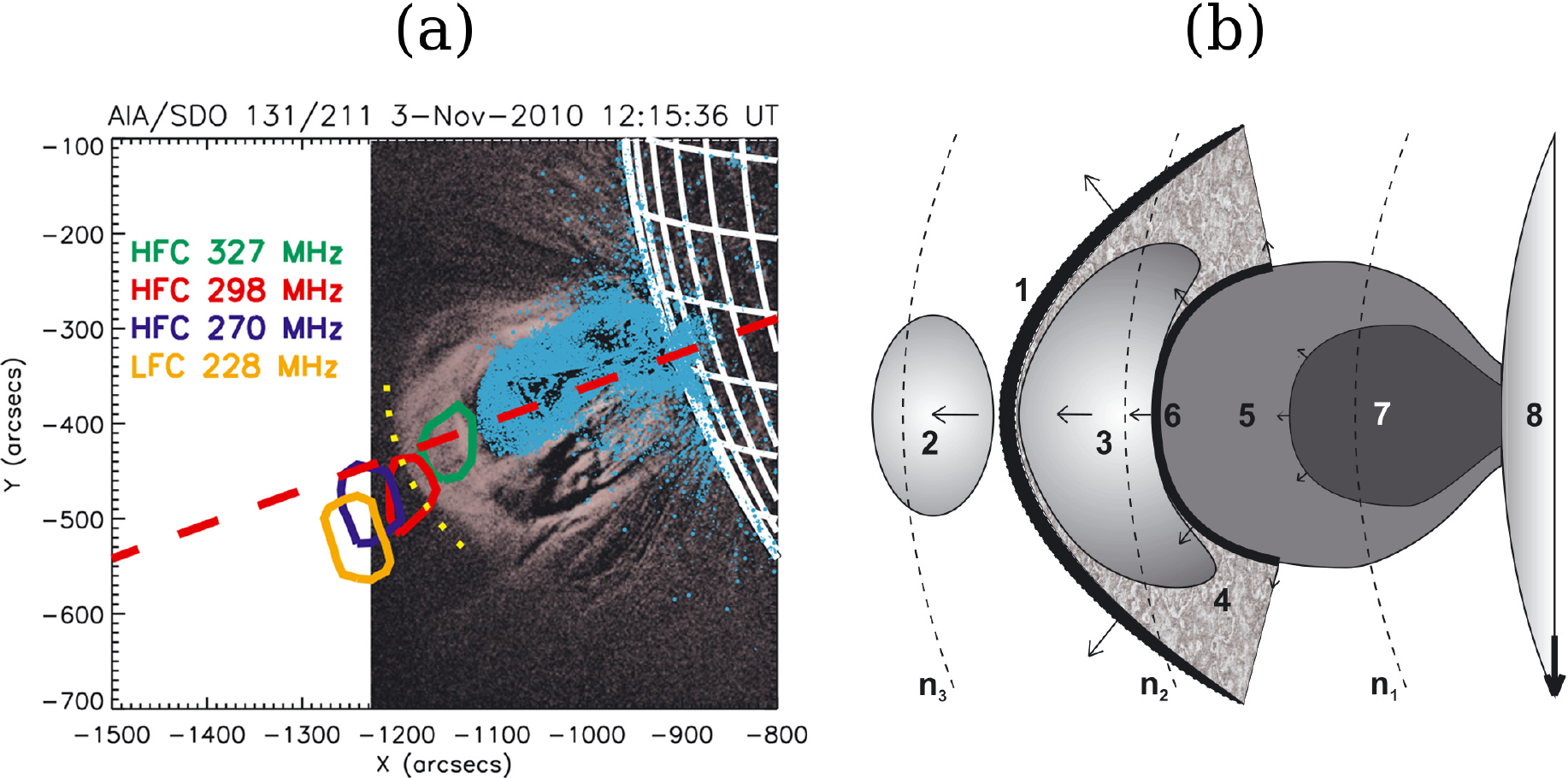}
        \caption{
        \textbf{Radio bursts, magnetic reconnection and turbulence ahead of the leading edge of an erupting solar magnetic flux rope}. Observation of a coronal mass ejection associated with an eruptive flare near the Sun's limb at 12:15:36 UT on 2010 November 3. Composite base-difference EUV images captured by SDO/AIA in 131 \r{A} (turquoise) and 211 \r{A} (purple). The yellow dotted parabola denotes the approximate front boundary layer of the erupting plasmas seen by AIA at 211 \r{A} ahead of a magnetic flux rope. Solid lines of various colours denote the NRH contours (95\% of the peak flux), marking the locations of centroids of the source regions of type-II radio bursts at different frequencies. The red dashed line denotes a projection of the radius-vector passing through the X-ray flare on the image plane. (b) Sketch of the eruptive event of (a) viewed from the heliographic north pole. The number notations denote, respectively, the: (1) hypothetical shock wave, (2) LFC of type-II radio bursts, (3) HFC of type-II radio bursts, (4) turbulent shock-sheath, (5) erupting warm (T $\sim$1-2 MK) plasma rim, (6) leading edge, (7) erupting magnetic flux rope with hot (T$\sim$10 MK) plasma, (8) photosphere. The black thick arrow denotes the direction to the Earth. The black thin arrows denote, respectively, the direction of: erupting magnetic flux rope, erupting warm plasma rim, shock wave, source motion of LFC/HFC sources. The lengths of arrows are proportional to the corresponding velocity of motions. The black dashed arc-lines denote the background electron plasma density: $n_1 > n_2 > n_3$. Magnetic reconnections can accelerate electrons at the shock-sheath and generate radio bursts via Langmuir turbulence.
        [reproduced from Ref. Zimovets et al. (2012)]}
        \label{fig:1}
    \end{figure}

    \subsubsection{Coronal mass ejection}
    A coronal mass ejection (CME) erupting from the solar limb was observed (Zimovets et al. \citeyear{zimovets2012spatially}) by ground and space instruments on 2010 November 3, following a C4.9 class flare peaking at 12:15:09 UT in active region AR 11121
    Figure \ref{fig:1}(a) shows a composite image at 12:15:36 UT of type-II radio bursts at four different frequencies detected by the Nan\c{c}ay Radioheliograph (NRH) and EUV viewing of the erupting multithermal plasmas detected at 131 \r{A} and 211 \r{A} by the Atmospheric Imaging Assembly (AIA) onboard the Solar Dynamics Observatory (SDO). 
    All four high-frequency component (HFC) and low-frequency component (LFC) radio sources, with frequencies decreasing upwards, are located above the leading edge (ejecta canopy, Foullon et al. \citeyear{foullon2011magnetic}) of the erupting plasma seen at 131 \r{A}. 
    This observation suggests that HFC (LFC) radio bursts are emitted at the downstream (upstream) regions of the shock, respectively, thus rendering support for the split-band model of type-II radio bursts proposed by Smerd et al. (\citeyear{smerd1975split}). 
    Figure 1(b) shows a sketch of different regions of this erupting CME. Coronal vortices driven by the magnetic Kelvin-Helmholtz instability were detected (Foullon et al. \citeyear{foullon2011magnetic}) at the leading edge of this erupting CME plasma seen at 131 \r{A}, which can induce magnetic reconnection. 
    Small-scale current sheets and magnet flux ropes embedded in the intermittent plasma turbulence at the downstream shock-sheath and the leading edge of ejecta can drive magnetic reconnection that accelerates electron beams, resulting in the generation of the HFC radio emissions via a beam-plasma instability and Langmuir turbulence (Chian and Alves \citeyear{chian1988nonlinear}).   
    The LFC source, situated in the upstream region, can be explained in the frame of some standard shock wave theories, e.g., the shock drift acceleration mechanism, and/or upstream beam-plasma instability and Langmuir turbulence.  \\
    \subsubsection{Interplanetary coronal mass ejection}
    \begin{figure}[ht]
        \centering
        \includegraphics[width=0.6\linewidth]{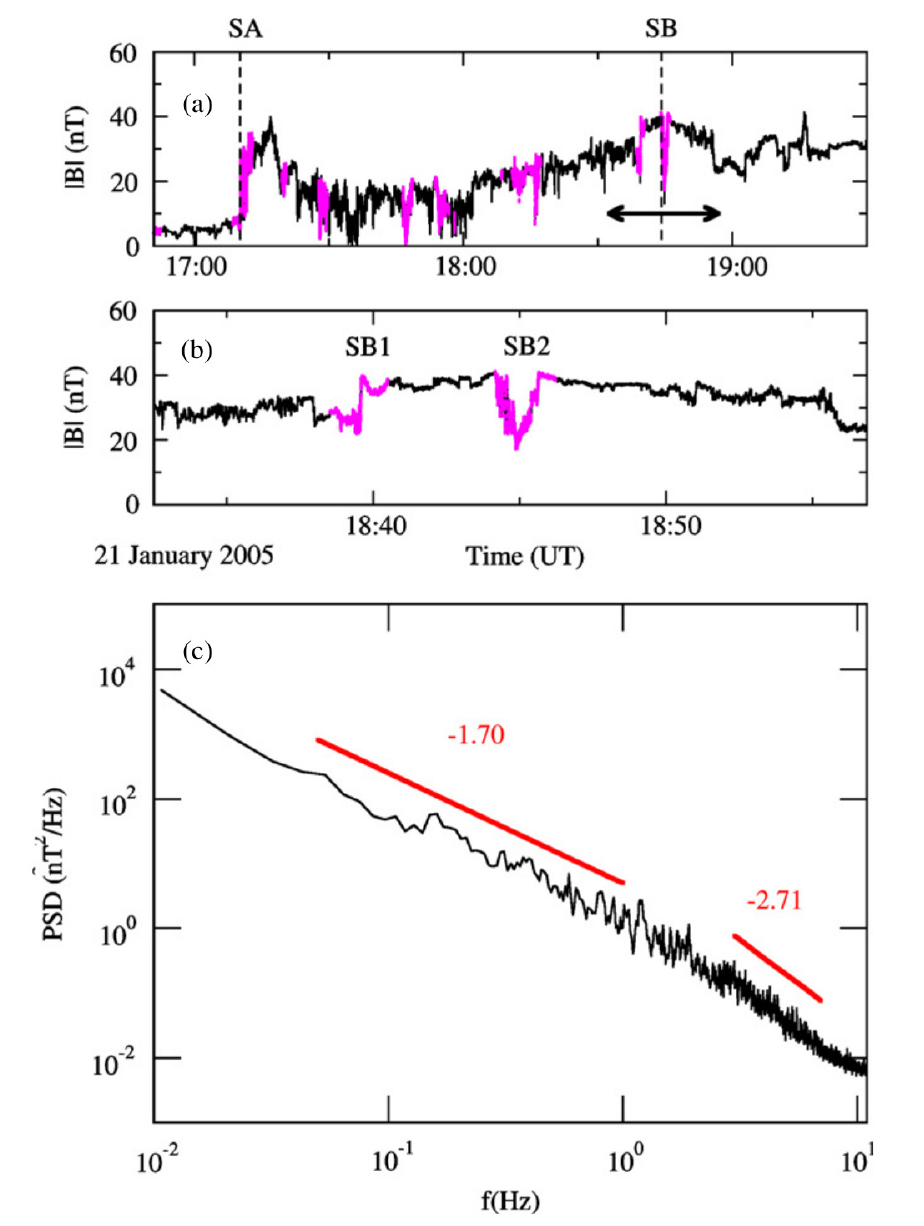}
        \caption{\textbf{Current sheets and turbulence at an interplanetary shock-sheath}.
        Observation of current sheets and magnetic turbulence by Cluster-1 at the turbulent shock-sheath (the time interval between SA and SB) of ICME of 2005 January 21. (a) Time series of $\abs{B}$(nT) superposed by small-scale current sheets found by the Li (\citeyear{li2008identifying}) method, for the critical angle $\theta = 60^\circ$  and the timescale $\tau= 120$ s. Magenta dots denote the points associated with a current sheet. SA indicates the shock arrival. (b) An enlarged view of the time interval at the leading edge of ejecta indicated by a bar in (a). $SB_{1}$ and $SB_{2}$ indicate the two current sheets detected at the leading edge (SB) of the ICME ejecta.  (c) Power spectral density (PSD, nT${}^2$ Hz${}^{-1}$) of $\abs{B}$ for the time interval (b) of magnetic intermittent turbulence; straight lines denote the inertial and dissipative subranges, respectively. The spectral indices are computed by a linear regression of the log-log PSD data.
        [reproduced from Ref. Chian and Mu\~noz (2011)]}
        \label{fig:2}
    \end{figure}
     \begin{figure}[ht]
        \centering
        \includegraphics[width=0.6\linewidth]{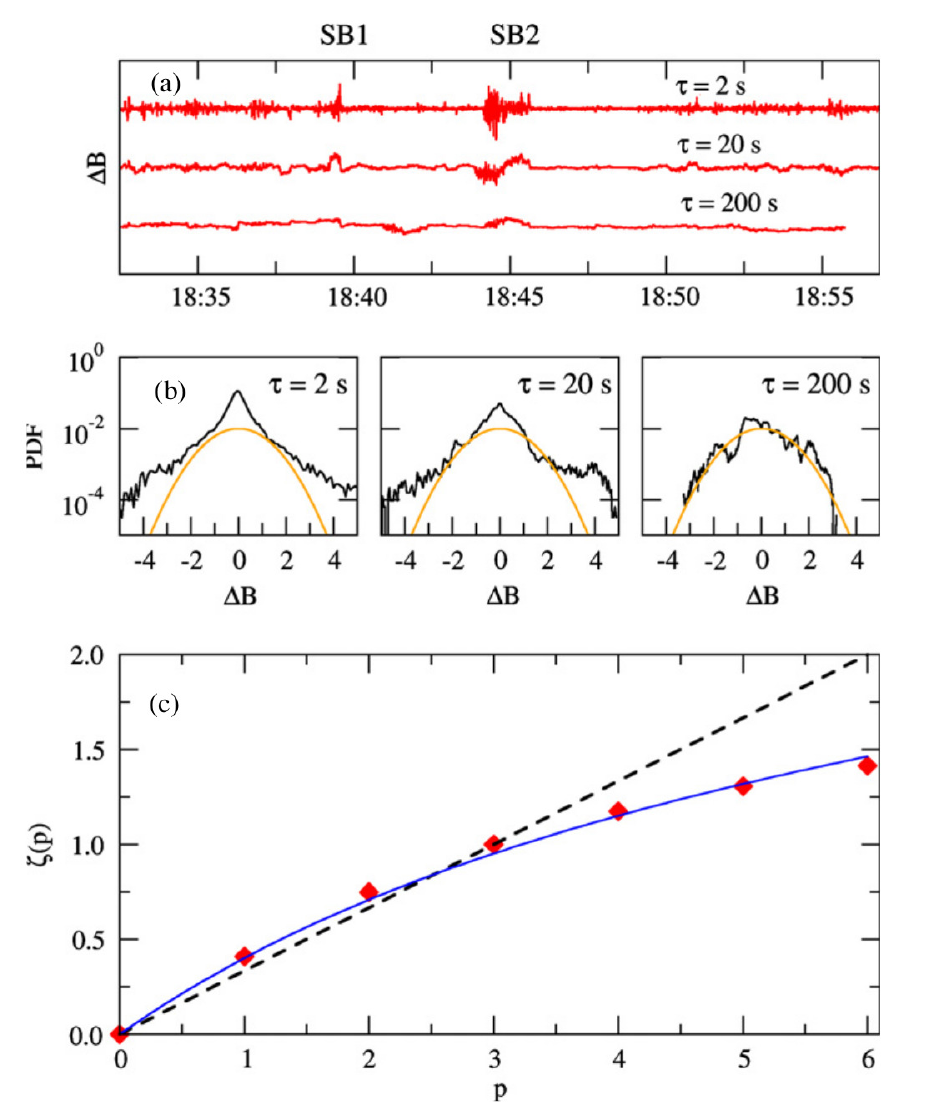}
        \caption{\textbf{Intermittency, non-Gaussianity, and multifractality at a shock-sheath: Interplanetary coronal mass ejection of 2005 January 21}. Scale dependence for three different timescales ($\tau = 2$ s, 20 s, and 200 s) of the time series of Figure 2(b). (a) The time series of two-point differences of normalized magnetic-field $\Delta B$, showing spiky bursts of intermittent structures at small timescales ($\tau = 2$ s and 20 s) at the vicinity of two current sheets $SB_1$ and $SB_2$.  (b) The probability density function (PDF) of $\Delta B$ superposed by a Gaussian PDF (orange line), showing non-Gaussian fat-tails at small-scales ($\tau =  2 $ s and 20 s). (c) Scaling exponent $\zeta$ of the pth-order structure function for magnetic fluctuations (red diamonds), superposed by the K41 self-similar scaling (black dashed line), and the multi-fractal prediction (She and Leveque \citeyear{she1994universal}, M\"uller and Biskamp \citeyear{muller2000scaling})  of the She-Leveque model of magnetic turbulence (blue curve). The departure of the scaling exponent from self-similarity is an indicative of multifractal turbulence due to its embedded intermittent structures. [reproduced from Ref. Chian and Mu\~noz (2011)]}
        \label{fig:3}
    \end{figure}
    \begin{figure}[ht]
        \centering
        \includegraphics[width=0.8\linewidth]{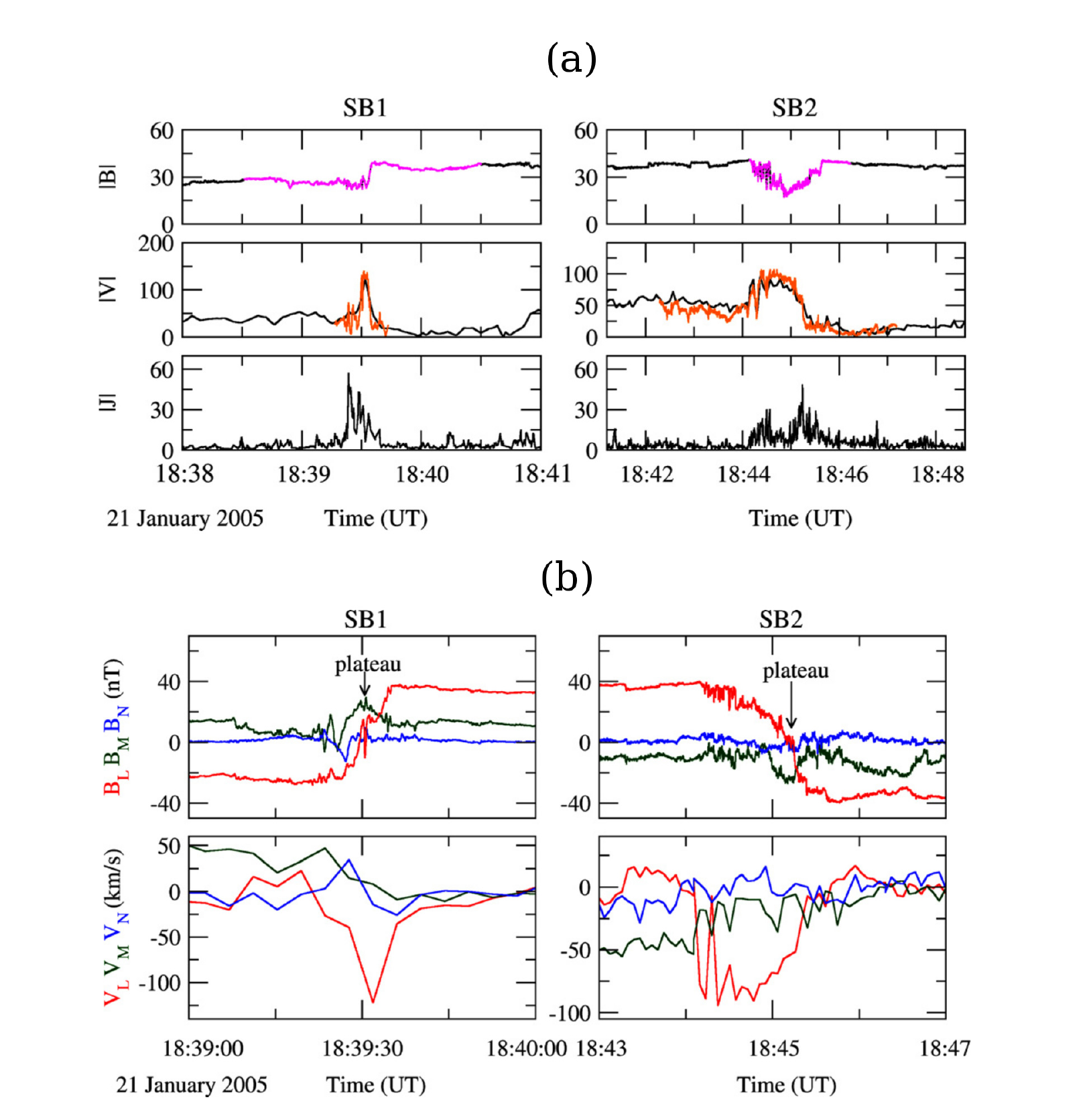}
        \caption{\textbf{Magnetic reconnection at the leading edge of a magnetic flux rope: Interplanetary coronal mass ejection of 2005 January 21}. Observation of magnetic reconnection related to the current sheets $SB_1$ and $SB_2$ (magenta) at the leading edge of ICME. (a) The modulus of magnetic field $\abs{B}$(nT) (an enlarged view of Figure 2(b)); the modulus of the plasma velocity $\abs{V}$ (km $s^{-1}$) (black) and the plasma velocity (orange) predicted by the magnetic reconnection theory of Sonnerup et al. (\citeyear{sonnerup1981evidence}); the modulus of the current density $\abs{J}$ (nA $m^{-2}$) calculated by the multi-spacecraft curlometer technique of Dunlop et al. (\citeyear{dunlop2002four}). (b) The components of magnetic field $B_L$ (red), $B_M$ (green), and $B_N$ (blue) in the LMN coordinates measured by Cluster-1; the components of the plasma velocity $V_L$ (red), $V_M$ (green), and $V_N$ (blue). The observational evidence of bifurcated current sheets $SB_1$ and $SB_2$ are given by a plateau at $B_L$ in the middle of each bifurcated current sheet,  parallel/anti-parallel Alfv\'en waves at two edges of the current sheets, and jets are the signatures of magnetic reconnection. [reproduced from Ref. Chian and Mu\~noz (2011)]}
        \label{fig:4}
    \end{figure}
    An interplanetary coronal mass ejection (ICME) was observed  (Foullon et al. \citeyear{foullon2007multi}) by multi-spacecraft at near-Earth solar wind and the Earth's magnetosheath on 2005 January 21-22.   The probable source of this ICME was a halo CME associated with an X7.1 flare from AR 10720. Figure 2(a) shows the time series of the modulus of magnetic field at the ICME turbulent shock-sheath measured by Cluster-1 (Chian and Mu\~noz \citeyear{chian2011detection}), where SA denotes the shock arrival, SB denotes the leading edge of ICME ejecta, and the interval between SA and SB denotes the turbulent shock-sheath embedded by a large number of small-scale current sheets and magnetic flux ropes. Two current sheets SB1 and SB2 are seen at an enlarged view given by Fig. 2(b) of the leading edge region marked by a horizontal bar in Figure 2(a). The power spectral density of Fig. 2(b) shows that the magnetic turbulence at the leading edge of ejecta is fully-developed, exhibiting a Kolmogorov spectral index of $-5/3$ in the inertial range and a spectral index of $-2.71$ at the dissipative range.  \\
    
    The normalized two-point differences of $\abs{B}$ of Fig. 2(b) for three different timescales ($\tau = 2$ s, 20 s, 200 s) shown in Figs. 3(a) demonstrate that the magnetic field fluctuations at the leading edge of ICME ejecta become more intermittent as the scale becomes smaller, evidenced by spiky bursts (intermittent structures) in the vicinity of two current sheets SB1 and SB2. Figure 3(b) shows that the corresponding PDF displays non-Gaussian fat-tails at small scales which is a signature of the intermittent structures. Moreover, Figure 3(c) shows that the scaling exponent $\zeta$ displays a noticeable departure from self-similarity and monofractality at higher orders of the structure function, indicative of a multifractal behavior.  The intermittency, non-Gaussianity, and multifractality of intermittent magnetic turbulence seen in Fig. 3 are the manifestation of coherent structures such as small-scale current sheets and magnetic flux ropes which can induce magnetic reconnection.\\
    
    The evidence of magnetic reconnection related to the two current sheets SB1 and SB2 is given in Fig. 4, which shows that the modulus of the ion velocity measured by Cluster-1 is close to the modulus of the reconnecting-jet velocity predicted by the magnetic reconnection theory of Sonnerup et al. (\citeyear{sonnerup1981evidence}). Localized large-amplitude current densities are seen at the locations of SB1 and SB2. The signature of bifurcated current sheets is seen in $\abs{J}$ and the plateau in $B_L$. In particular, Fig. 4 shows that $V_L$ is anti-correlated (correlated) with $B_L$ at the leading (trailing) boundary of SB1, and $V_L$ is correlated (anti-correlated) with $B_L$ at the leading (trailing) boundary of SB2. This provides the evidence of Alfv{\'e}n waves propagating parallel/anti-parallel to the ambient magnetic field,  associated with magnetic reconnection exhausts (Wei et al. \citeyear{wei2003magnetic}; Gosling et al. \citeyear{gosling2005direct}; Phan et al. \citeyear{phan2006magnetic}).
    
    \subsection{Interplanetary rope-rope magnetic reconnection}
    \subsubsection{ICME-ICME merger}
     \begin{figure}[htp]
        \centering
        \includegraphics[width=0.8\linewidth]{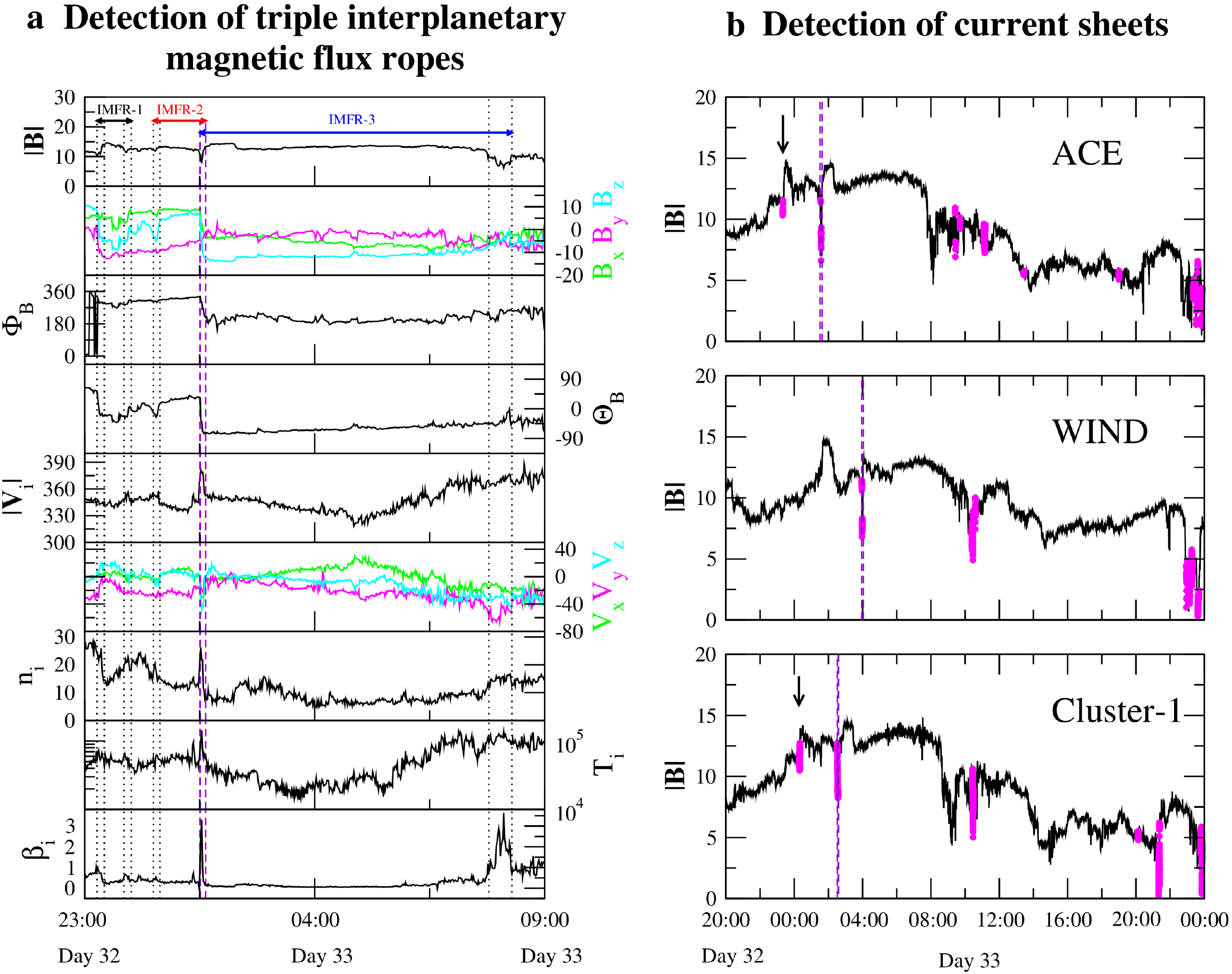}
        \caption{\textbf{Interplanetary rope-rope magnetic reconnection: Merger of multiple coronal mass ejections}. Observation of multiple interplanetary magnetic flux ropes and current sheets on 2002 
        February 1-2. (a) Time series of the ACE magnetic field and plasma parameters. The horizontal bar marks the interval of each magnetic flux rope. The vertical dotted lines mark the front and rear boundary layers of triple-IMFR, respectively. From top to bottom: the modulus of magnetic field  $\abs{B}$(nT); three components of $B$(nT) in the GSE coordinates; azimuth angle $\Phi_{B}$ 
        (degrees); latitude angle $\Theta$ (degrees); modulus of ion bulk velocity $\abs{V_{i} }$ (km $s^{-1}$ ); three components of ion bulk velocity $V_{i}$ (km $s^{-1}$) in the GSE coordinates, where $V_x$ has been shifted by $+350$ km $s^{-1}$ ; ion number density $n_{i} \, (cm^{-3} )$; 
        ion temperature $T_{i}$ (eV); and ion plasma beta $\beta_{i}$. 
        (b) Current sheets (magenta dots) detected by the method of magnetic shear angle, superposed on the time series of $\abs{B}$ of ACE, Wind, and Cluster-1. Two purple vertical dashed lines mark the site of rope-rope magnetic reconnection at the interface region of IMFR-2 and IMFR-3, where a spike in $\abs{V_{i}} $, $n_{i}$, $T_{ i}$, and $\beta_{bi}$ are seen in Figure 5(a) and a current sheet is seen in Figure 5(b). In addition, a current sheet is seen at the front boundary 
        layers of IMFR-1 (marked by arrow) by ACE and Cluster-1. [reproduced from Ref. Chian et al. (2016)]}
        \label{fig:5}
    \end{figure}
     \begin{figure}[ht]
        \centering
        \includegraphics[width=0.7\linewidth]{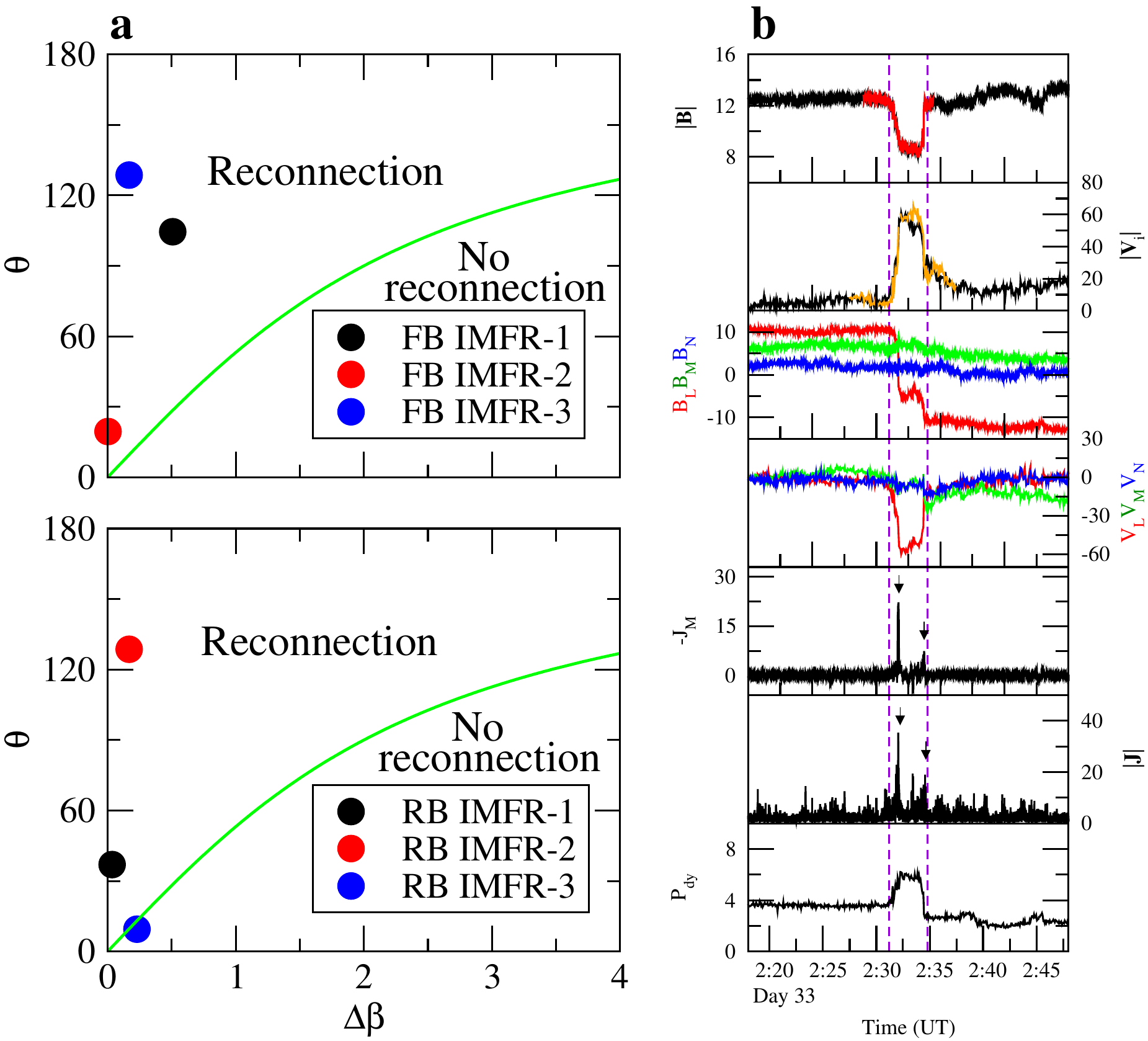}
        \caption{\textbf{Condition for magnetic reconnection at the interface region of two magnetic flux ropes: Interplanetary rope-rope magnetic reconnection of 2002 February 1-2}. (a) Condition for magnetic reconnection at the front-FB (rear-RB) boundary layers of triple-IMFR measured by ACE. (b) Signatures of rope-rope magnetic reconnection detected by Cluster-1 at the interface region of IMFR-2 and IMFR-2 marked by two purple vertical dashed lines of Figure 5. From top to bottom: $\abs{B}$(nT) is the modulus of magnetic field (black) and the associated current sheet (magenta dots); $\abs{V_i}$(km $s^{-1}$ ) is the modulus of the ion bulk velocity (black) and the velocity (orange) predicted by the magnetic reconnection theory; $B_L$ (red), $B_M$ (green), and $B_N$ (blue) are the components of the magnetic field B in the LMN coordinates, shear angle = $122^\circ$; $V_L$ (red), $V_M$ (green), and $V_N$ (blue) are the components of the ion bulk velocity $V_i$, where for visualization we have shifted the plasma velocities by the average solar wind velocity; $J_M$ (nA$m^{-2}$) is calculated from $B_L$; $\abs{J}$ (nA$m^{-2}$ ) is the modulus of current density calculated by the multi-spacecraft curlometer technique based on B data of four Cluster spacecraft; $n_i (cm^{-3 })$ is the ion number density; and the dynamic pressure is $P_{dy}$ (nPa). The plots of $J_M$ and $\abs{J}$ show double peaks (marked by arrows) at the leading (trailing) edges of the bifurcated current sheet, where correlated (anti-correlated) $B_L$ and $V_L$ are evidence of parallel (anti-parallel) Alfv\'en waves. [reproduced from Ref. Chian et al. (2016)]}
        \label{fig:6}
    \end{figure}
     \begin{figure}[ht]
        \centering
        \includegraphics[width=0.8\linewidth]{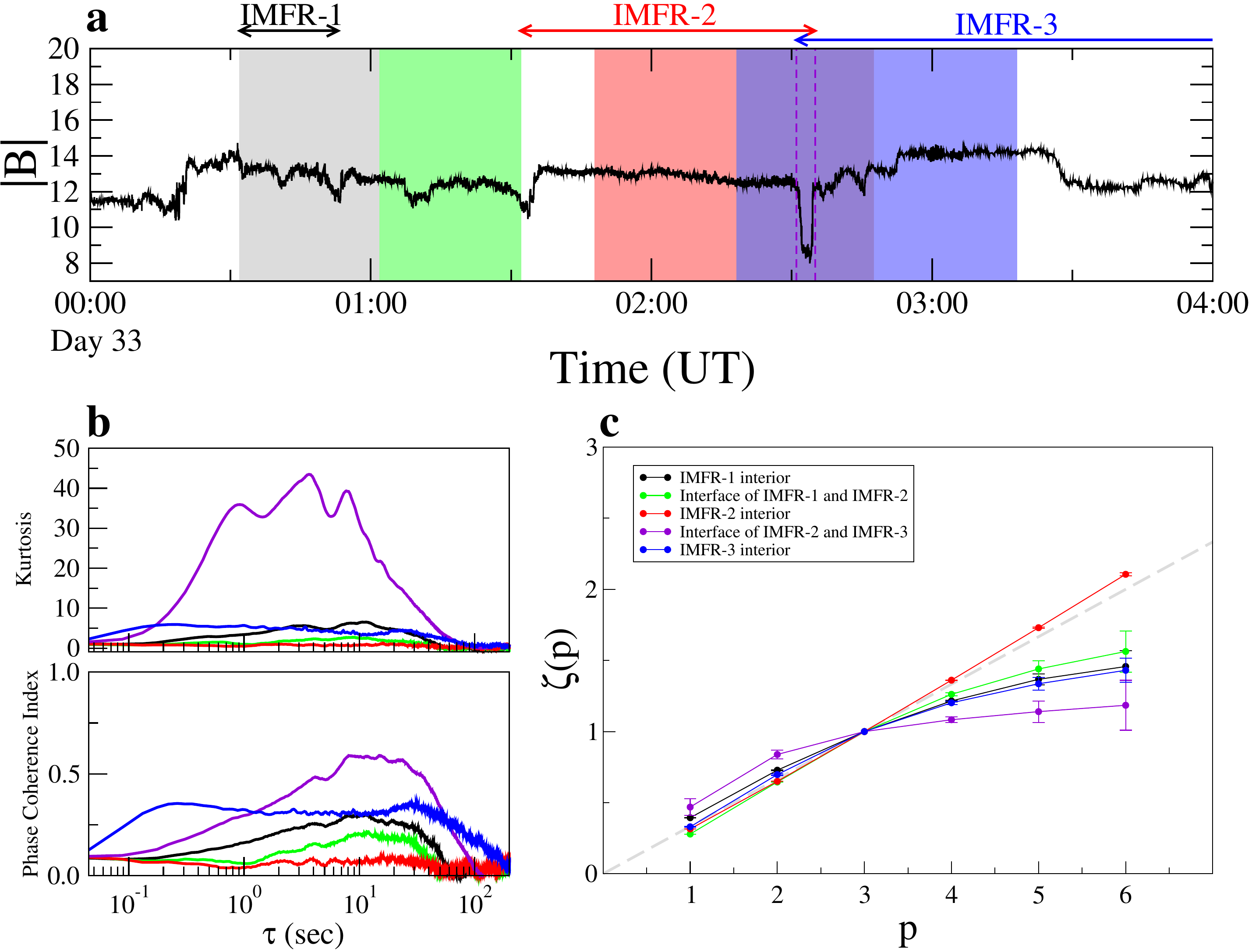}
        \caption{\textbf{Genesis of intermittent turbulence: Interplanetary rope-rope magnetic reconnection of 2002 February 1-2. Observation of interplanetary intermittent magnetic turbulence by Cluster-1 on 2002 February 2}. (a) Time series of the modulus of magnetic field $\abs{B}$(nT) detected by Cluster-1 divided into five regions of 30 minutes each: IMFR-1 interior (black), interface region of IMFR-1 and IMFR-2 (green), IMFR-2 interior (red), interface region of IMFR-2 and IMFR-3 (purple), and IMFR-3 interior (blue); the interval of each IMFR is denoted by a horizontal bar. The two purple vertical dashed lines indicate the interface region of IMFR-2 and IMFR-3. (b) Quantification of amplitude-phase synchronization by kurtosis and phase coherence index of $\abs{B}$ as a function of the timescale $\tau$. (c) Scaling exponents $\zeta(p)$ of the pth-order structure function of two-point differences of B (with error bars), superposed by the K41 self-similar scaling (gray dashed line). [reproduced from Ref. Chian et al. (2016)]}
        \label{fig:7}
    \end{figure}
    Multiple coronal mass ejections can merge with each other in the solar atmosphere and solar wind. For example, Kozyra et al. (\citeyear{kozyra2013earth}) detected two interacting CMEs behind the ICME shock-sheath of 2005 January 21-22 discussed in Section 3.1.2. Chian at al. (\citeyear{chian2016genesis}) used multi-spacecraft solar wind data at the Lagrangian point L1 and the Earth's foreshock on 2002 February 1-2 to investigate the interaction of three interplanetary magnetic flux ropes (IMFRs), and show that the magnetic reconnection exhaust event studied by Phan et al. (\citeyear{phan2006magnetic}) takes place at the interface region between IMFR-2 and IMFR-3 shown in Fig. 5.
     Figure 5(a) gives an overview of \textit{ACE} plasma data where the front and rear boundary layers of three IMFRs are identified. Figure 5(b) shows the small-scale current sheets detected by \textit{Wind}, and \textit{Cluster-1} using the method of magnetic shear angle (Li \citeyear{li2008identifying}). The \textit{ACE} suprathermal electron pitch angle spectrograms present evidence of accelerated energetic electrons at the site of ICME-ICME merger where an intense small-scale current sheet is found. Magnetic flux ropes are bundles of helical, current-carrying, magnetic field lines writhing about each other and spiralling around a common axis as discused in Section 2. The structures of IMFR-1 and IMFR-2 can be reconstructed by the Grad-Shafranov method (Hu et al. \citeyear{hu2004multiple}; Chian et al. \citeyear{chian2016genesis}); the reconstruction of IMFR-3 is difficult since the spacecraft path is far away from the center of the magnetic flux rope.\\
     
    Interplanetary magnetic flux ropes can erode a substantial amount of outer magnetic flux via magnetic reconnection at their boundary layers as they propagate away from the Sun through the heliosphere. In fact, the boundary layers of IMFRs can be recognized (Wei et al. \citeyear{wei2003magnetic}; Chian et al. \citeyear{chian2016genesis}) by noting the plasma characteristics of prior or ongoing magnetic reconnection with enhanced variations in the magnetic field strength, plasma speed, plasma density, plasma temperature, and ion plasma $\beta_i$, as seen in Figs. 4 and 5. The following condition for magnetic reconnection was derived by Swisdak et al. (\citeyear{swisdak2010vector})\\
    
    \begin{equation}
       {\Delta} \beta < 2(L_p/di) \tan(\theta/2),
    \end{equation}
    
     \noindent
     where $\Delta \beta$ denotes the jump in plasma $\beta$ across a boundary layer, $\theta$ denotes the shear angle between the reconnecting fields, $L_p$ denotes a typical pressure scale length near the X-line, and $di$ denotes the ion skin depth.  Figure 6(a) illustrates the condition for magnetic reconnection using the \textit{ACE} solar wind data of 2002 February 1-2, superposed by the values of ($\theta$, $\Delta \beta$) of the front and rear boundary layers of triple-IMFR, which shows that during this event the interface region of IMFR-2 and IMFR-3 is the most likely site for magnetic reconnection. Figure 6(b) provides an overview of \textit{Cluster-1} observational evidence of rope-rope magnetic reconnection at this event. Note that the large-amplitude dynamic pressure pulse $P_{dy}$ driven by the interplanetary magnetic reconnection, seen in Fig. 6(b), can compress the magnetosphere, raise the dayside magnetospheric magnetic field strength, and initiate resonant magnetic field perturbations in high-latitude ground magnetometers (Sibeck et al. \citeyear{sibeck1989solar}).\\
     
    Solar wind intermittent turbulence, consisted of nonlinear Alfv\'en fluctuations and coherent structures such as magnetic flux ropes and current sheets, is a driver of geomagnetic and auroral activities (D'Amicis et al. \citeyear{d2010geomagnetic}, \citeyear{d2020effect}). A geomagnetic storm on 2002 February 2 was driven by the long-duration negative $B_z$ in the IMFR-3 seen in Fig. 5(a). The onset of this storm was triggered by the dynamic pressure pulse seen in Fig. 6(b). Hence, the study of the genesis of solar wind intermittent turbulence is crucial for space weather forecasting. The degree of intermittency in turbulence can be quantified by kurtosis and phase coherence index (Hada et al. \citeyear{hada2003phase}) of the fluctuations, which measure the degree of amplitude and phase synchronization in multiscale interactions (Chian et al. \citeyear{chian2016genesis}), respectively. Moreover, the degree of multifractality in turbulence can be quantified by the scaling exponent $\zeta(p)$ of pth-order structure functions. A comparative study of five regions of the triple-IMFR event of 2002 February 2 using \textit{Cluster-1} $\abs{B}$ data demonstrates in Fig. 7 that the degree of intermittency and multifractality is highest at the interface region of IMFR-2 and IMFR-3. Hence, the interplanetary rope-rope magnetic reconnection related to ICME-ICME merger provides a key source for the genesis of solar wind intermittent turbulence.\\
    \subsubsection{Kurtosis-skewness relation} 
     \begin{figure}[ht]
        \centering
        \includegraphics[width=0.8\linewidth]{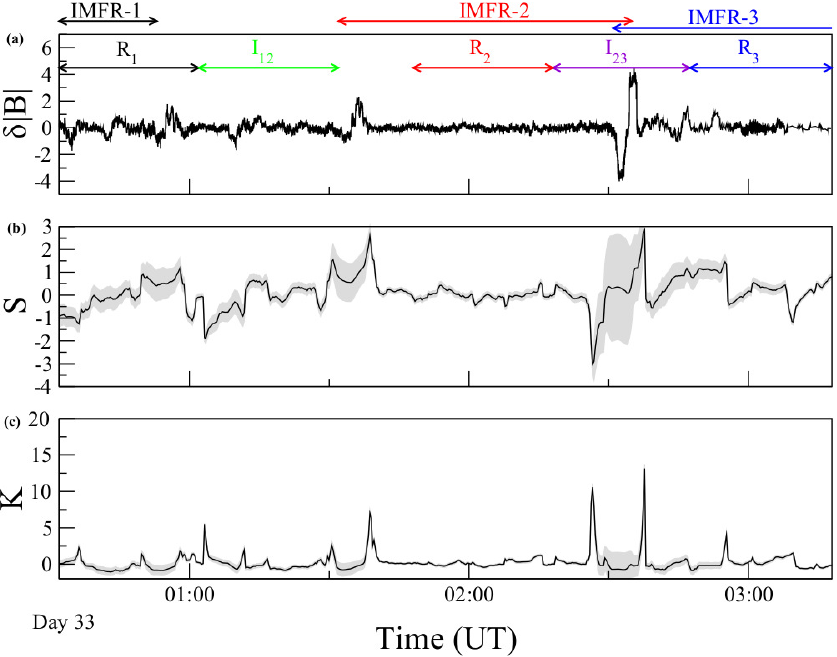}
        \caption{\textbf{Times series of kurtosis and skewness of intermittent turbulence: Interplanetary rope-rope magnetic reconnection of 2002 February 1-2}. (a) Time series of $\delta B$ ($\tau= 100$ s) from 00:00 to 04:00 UT on 2 February 2002. Five regions of 30 min each are marked by different colours: interior of IMFR-1 (R1, black), interface region of IMFR-1 and IMFR-2 ($I_{12}$, green), the interior of IMFR-2 (R2, red), the interface of IMFR-2 and IMFR-3 ($I_{23}$, violet), and the interior of IMFR-3 (R3, blue). Time series of skewness S (b) and kurtosis K (c) calculated using a sliding overlapping window. The gray area represents the standard deviation calculated in each window. The bifurcated current sheets associated with rope-rope magnetic reconnection at the interface region of IMFR-2 and IMFR-3 are detected by the two sharp spikes in $\abs{\delta B}$, S, and K. 
        [reproduced from Ref. Miranda et al. (2018)]}
        \label{fig:8}
    \end{figure}
     \begin{figure}[ht]
        \centering
        \includegraphics[width=0.7\linewidth]{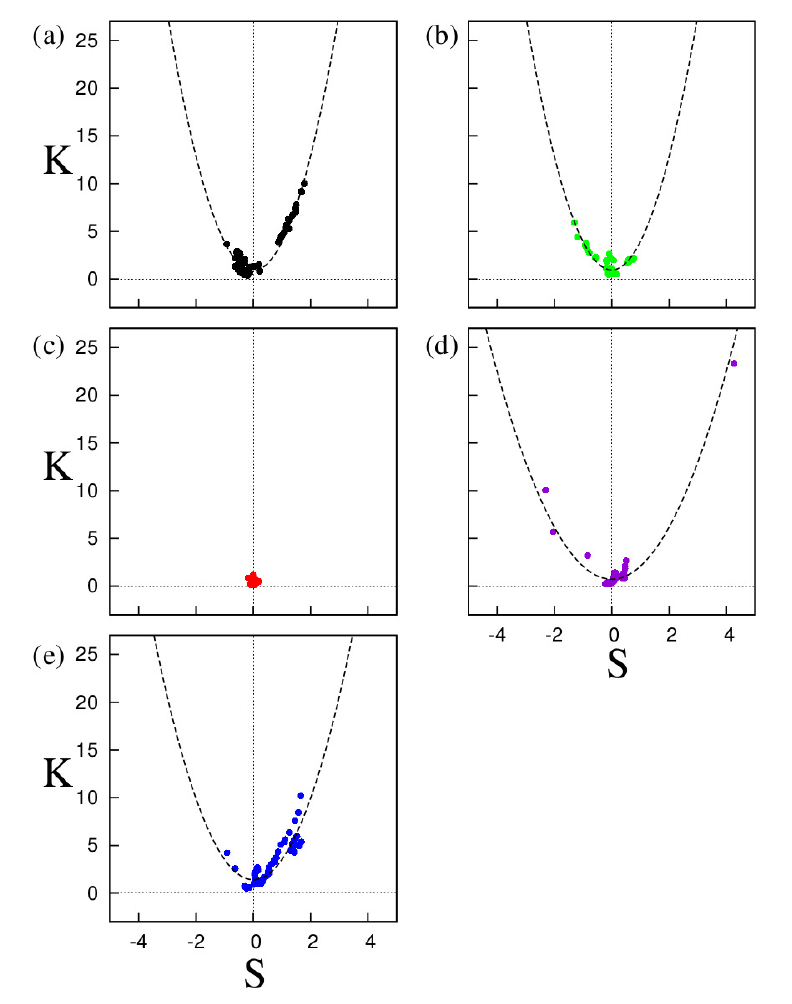}
        \caption{\textbf{Kurtosis-skewness relation of intermittent turbulence: Interplanetary rope-rope magnetic reconnection of 2002 February 1-2}. Kurtosis K as a function of skewness S calculated from the time series of $\abs{\delta B(\tau)})$, where $\tau = 10$ s, for five regions marked in Figure 8: (a) the interior of IMFR-1, (b) the interface of IMFR-1 and IMFR-2, (c) the interior of IMFR-2, (d) the interface of IMFR-2 and IMFR-3, and (e) the interior of IMFR-3. The dashed line in each panel indicates the least-square fit with the parabolic function $K = \alpha S^{2} + \beta$.
         [reproduced from Ref. Miranda et al. (2018)]}
        \label{fig:9}
    \end{figure}
    A unique parabolic relation linking skewness to kurtosis, i.e., third- and fourth-order structure functions, respectively, have been observed in solar wind turbulence near interplanetary shocks (V\"or\"os et al. \citeyear{voros2006cross}) and drift-interchange turbulence in a toroidal plasma device (Labit et al. \citeyear{labit2007universal}). This relation can elucidate the universal statistical properties of intermittent turbulence. Miranda et al. (\citeyear{miranda2018non}) investigated the kurtosis-skewness relation of the triple interplanetary magnetic flux rope event studied by Chian  et al. (\citeyear{chian2016genesis}). Figure 8 shows the time series of two-point differences ($\tau = 100 s$) of the modulus of magnetic field measured by \textit{Cluster-1} and the corresponding skewness and kurtosis, where the five regions of 30 min each are the same as specified in Fig. 7(a). It is evident from Fig. 8 that the interface region of IMFR-2 and IMFR3, where rope-rope magnetic reconnection occurs, is readily identified by the sharp spike in $\delta \abs{B}$, skewness, and kurtosis. \\
    
    The plot of kurtosis as a function of skewness ($\tau= 10$  s) for five regions of Figs. 7 and 8 is shown in Fig. 9. With the exception of the region $R_2$ in the interior of IMFR-2 (Fig. 9(c)) where the magnetic fluctuations are nearly Gaussian as seen in Fig. 7, the kurtosis-skewness relation for all other regions show a parabolic shape, confirming the non-Gaussian, intermittent and multifractal nature of magnetic fluctuations in these regions as seen in Fig. 7. The functional relation between kurtosis and skewness for each region can be described by the quadratic equation
    \begin{equation} \label{eq:2}
        K=\alpha\, S^2 + \beta,
    \end{equation}
    \noindent
    where $\alpha$ and $\beta$ are the coefficients that characterize a parabolic curve which can be computed by applying a least-square fit between the values of kurtosis and skewness obtained from the observational data and Eq. (\ref{eq:2}), following the Levenberg-Marquardt algorithm (Miranda et al. \citeyear{miranda2018non}). A correlation index $r$ can be computed to measure the degree of correlation between the value of observed $K$ and the value of $K$ obtained empirically from Eq. (\ref{eq:2}).  Figure 9 shows that the parabolic kurtosis-skewness relation of the magnetic intermittent turbulence is enhanced (with $r=0.99$) at the interface region of IMFR-2 and IMFR-3 (Fig. 9(e) with $\alpha$ = 1.36 , $\beta= 0.72$) due to nonlinear Alfv\'en waves and coherent structures such as small-scale current sheets and magnetic flux ropes related to the interplanetary rope-rope magnetic reconnection discussed in Section 3.2.1. The studies of the kurtosis-skewness relation of V\"or\"os et al. (\citeyear{voros2006cross}) and Miranda et al. (\citeyear{miranda2018non}) show that the solar wind intermittent turbulence can be induced by cross-scale interactions.\\
    
    \subsubsection{Complexity-entropy relation} 
    
     \begin{figure}[ht]
        \centering
        \includegraphics[width=0.8\linewidth]{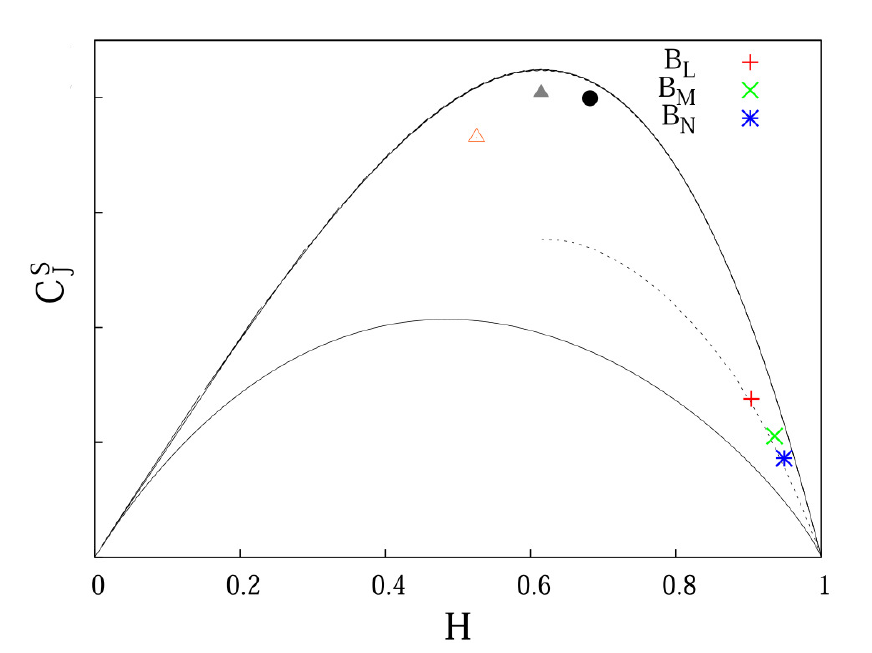}
        \caption{\textbf{Complexity-entropy relation of intermittent turbulence: Interplanetary rope-rope magnetic reconnection of 2002 February 1-2}. The complexity-entropy relation for $B_L$ (red plus), $B_M$ (green cross), and $B_N$ (blue asterisk) components of the \textit{Cluster-1} magnetic field during the reconnection exhaust event of the interplanetary rope-rope magnetic reconnection on 2002 February 2. The full black circle, open red triangle, and full gray triangle denote the chaotic time series of the logistic map, the skew tent map, and the H\'enon map, respectively, using the same parameter values of the chaotic maps as Rosso et al. (\citeyear{rosso2007distinguishing}) and Weck et al. (\citeyear{weck2015permutation}). The dotted line denotes stochastic fractional Brownian motions (fBm). This curve was computed by generating a time series of fBm with a Hurst exponent varying within the interval [0.025, 0.925] (Magg and Morales \citeyear{maggs2013permutation}). The crescent-shaped curves denote the maximum and minimum values of the complexity $C_J^S$ for a given value of the normalized Shannon entropy $H$, respectively. [reproduced from Ref. Miranda et al. (2021)]}
        \label{fig:10}
    \end{figure}
    The link between chaos, complexity, and intermittent turbulence can be clarified by computing the Jensen-Shannon complexity-entropy index, which is a statistical tool capable of distinguishing noise from chaos (Bandt and Pompe \citeyear{bandt2002permutation}; Rosso et al. \citeyear{rosso2007distinguishing}) and has been applied to characterize  chaos in laboratory plasmas (Magg and Morales \citeyear{maggs2013permutation}; Gekelman et al. \citeyear{ gekelman2014chaos}; Onchi et al. \citeyear{onchi2017permutation}; Zhu et al. \citeyear{ zhu2017chaotic}) and stochasticity in space plasmas (Weck et al. \citeyear{weck2015permutation}; Osmane et al.  \citeyear{osmane2019jensen}; Weygand and Kevilson \citeyear{ weygand2019jensen}; Good et al. \citeyear{good2020radial}). Miranda et al (\citeyear{miranda2021complexity}) used this tool to study the complexity-entropy of the LMN components of magnetic field of four magnetic reconnection exhaust events in solar wind. Figure 10 shows the results obtained by Miranda et al. (\citeyear{miranda2021complexity}) for the interplanetary rope-rope magnetic reconnection event observed by \textit{Cluster-1} on 2002 February 2 (Chian et al. \citeyear{chian2016genesis}). The complexity-entropy plane can be separated into three regions: 1) a low-entropy and low-complexity region corresponding to highly predictable systems (e.g., periodic attractor), 2) an intermediate-entropy and high-complexity region corresponding to unpredictable systems (e.g., chaotic attractor), and 3) a high-entropy and low-complexity region corresponding to stochastic systems (e.g., stochastic attractor) (Rosso et al. \citeyear{rosso2007distinguishing}). For the sake of comparison, some examples of chaotic and stochastic time series are given in Fig. 10 to mark the regions 2 and 3, respectively.   Figure 10 shows that the ($C_J^S$, $H$) values of all the LMN components of $B$ lie close to the bottom-right region of the complexity-entropy plane, indicating that the interplanetary magnetic fluctuations in this event are stochastic in nature, confirming the results of interplanetary turbulence obtained by Weck et al. (\citeyear{weck2015permutation}), Weygand and Kivelson (\citeyear{weygand2019jensen}), and Good et al. (\citeyear{good2020radial}). The $B_L$ component displays a lower degree of entropy and higher degree of complexity than the $B_M$ component, which in turn displays a lower degree of entropy and a higher degree of complexity than the $B_N$ component. All four events studied by Miranda et al. (\citeyear{miranda2021complexity}) obtained this same result. One event observed by \textit{Wind} on 1997 December 30 has a long exhaust duration with sufficient data points to determine the universal scaling exponent $\zeta(p)$ of magnetic fluctuations, which shows that $B_L$ is more intermittent (multifractal) than $B_M$, which in turn is more intermittent (multifractal) than $B_N$. Hence, Miranda et al. (\citeyear{miranda2021complexity}) concluded for this event that a higher degree of intermittency (multifractality) is related to a lower degree of the normalized Shannon entropy (H) and a higher degree of the Jensen-Shannon complexity ($C_J^S$) in the inertial range of interplanetary magnetic intermittent turbulence at the magnetic reconnection exhaust. \\
    \subsection{Earth's bow shock}
    The scenario of erupting CME in solar corona sketched by Fig. 1(b), consisting of CME shock-CME turbulent sheath-Leading edge of CME, is similar to another key region of the Sun-Earth system consisting of the Earth's bow shock-Turbulent magnetosheath-Magnetopause (Foullon et al. \citeyear{foullon2011magnetic}), where vortices driven by the magnetic Kelvin-Helmholtz instability at the Earth's dayside magnetopause can induce magnetic flux ropes and magnetic reconnections that mediate the solar wind-magnetosphere coupling (Kieokaew et al. \citeyear{kieokaew2020magnetic}). Hence, solar wind turbulence near the Earth's bow shock plays a key role in the transport of solar wind plasma to the Earth's magnetosphere and ionosphere.\\
    \subsubsection{Far and near upstream of the Earth's bow shock}
     \begin{figure}[ht]
        \centering
        \includegraphics[width=0.8\linewidth]{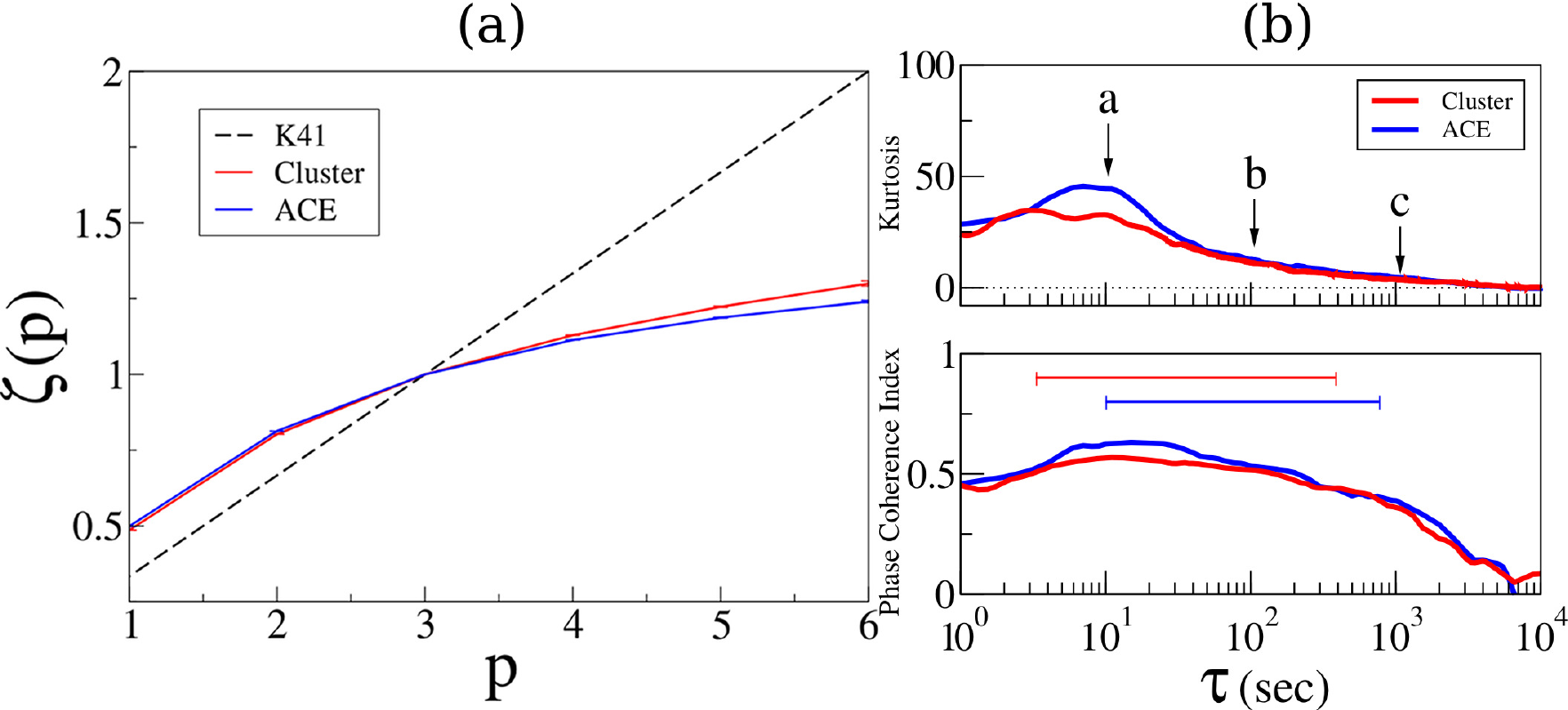}
        \caption{\textbf{Amplitude-phase synchronization of intermittent turbulence: A comparison of L1 and the Earth's foreshock}. (a) Scaling exponent $\zeta$ of the pth-order structure function determined by ESS fitting for Cluster-1 and ACE magnetic field fluctuations. The dashed line corresponds to K41 (self-similar) Kolmogorov scaling. (b)  Quantification of amplitude-phase synchronization by kurtosis and phase coherence index of $\abs{B}$ measured by Cluster-1 (red) and ACE (blue). Letters a, b and c denote scales $\tau = 10$, 100 and 1000 s, respectively. The bars denote the inertial subrange. The inverse of the ion cyclotron frequency $f_{ci}\sim 0.12\, Hz$ in the solar wind frame is $\tau \sim 8.3$ s, which is near the peak regions of kurtosis and phase coherence index.  [reproduced from Ref. Chian and Miranda (2009)]}
        \label{fig:11}
    \end{figure}
    Chian and Miranda (\citeyear{chian2009cluster}) carried out a comparative study of the degree of intermittency (multifractality) and amplitude-phase synchronization of solar wind turbulence far and near upstream of the Earth's bow shock. This analysis is based on the magnetic-field data measured simultaneously from 1 to 3 February 2002 by \textit{ACE} and \textit{Cluster-1}, respectively, during the time interval between the crossing of  \textit{Cluster-1} departing from the quasi-perpendicular shock to the crossing of  \textit{Cluster-1}  entering into the quasi-parallel shock. Note that the magnetic reconnection exhaust event studied by Phan et al. (\citeyear{phan2006magnetic})  and the triple interplanetary magnetic flux rope event studied by Chian et al. (\citeyear{chian2016genesis}) occur within this time interval. The computed scaling exponent $\zeta(p)$ in Fig. 11(a) shows that the degree of intermittency (multifractality) of the solar wind turbulence far upstream of the Earth's bow shock measured by \textit{ACE} at the Lagrangian point L1 is higher than near upstream of the Earth's bow shock measured by B. This result is confirmed by kurtosis and phase coherence index in Fig. 11(b), which demonstrates that the degree of amplitude-phase synchronization in multiscale interactions measured by \textit{ACE} is higher than \textit{Cluster-1}.  During this time interval \textit{Cluster-1} is located at the Earth's foreshock where the solar wind ions reflected from the Earth's bow shock can intensify the dissipation of interplanetary Alfv\'en waves via ion-cyclotron damping and other kinetic effects (Howes et al. \citeyear{howes2008kinetic}), leading to a decrease of amplitude-phase synchronization in multi-scale interactions. \\
     \subsubsection{Upstream and downstream of the Earth's bow shock}
      \begin{figure}[ht]
        \centering
        \includegraphics[width=0.5\linewidth]{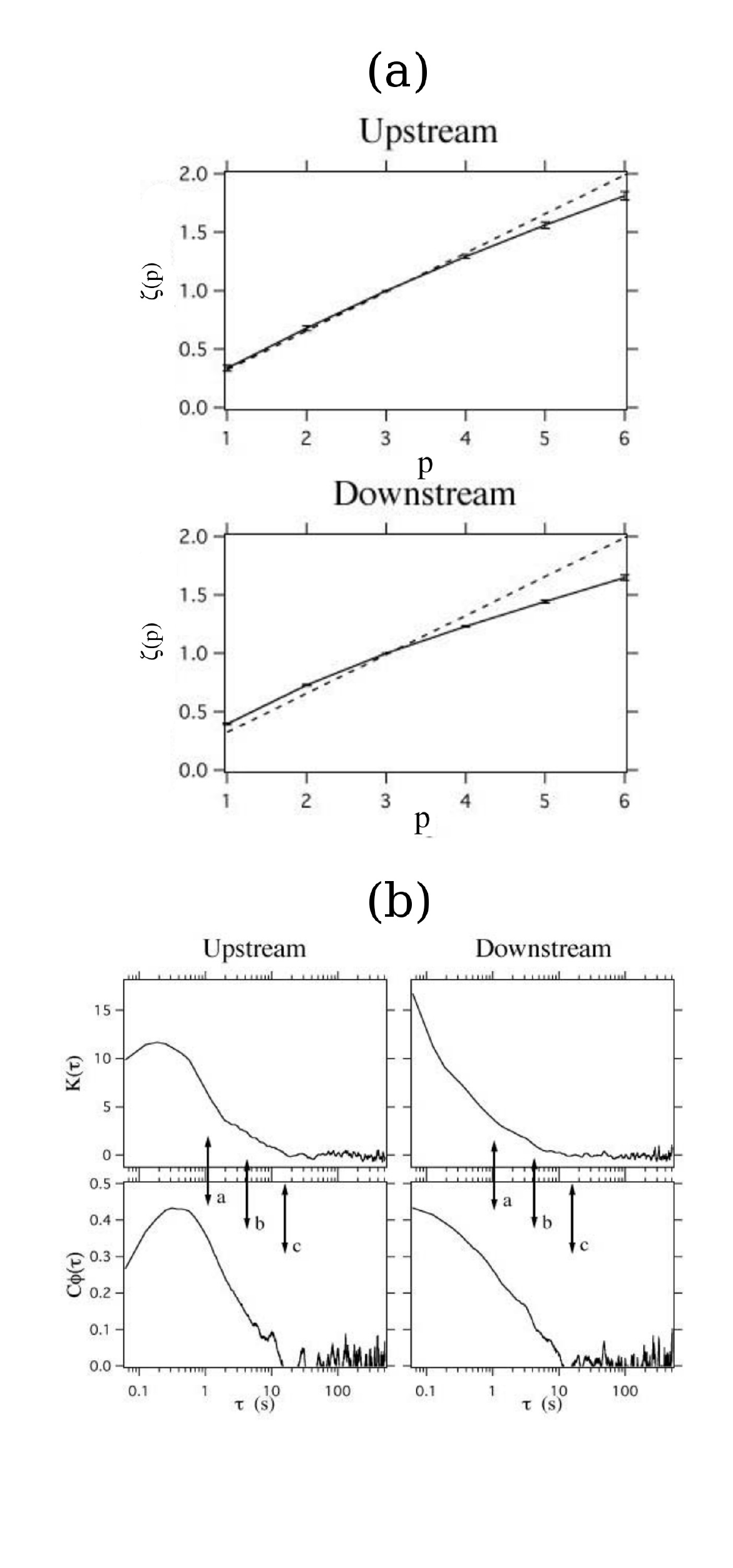}
        \caption{\textbf{Amplitude-phase synchronization of intermittent turbulence at the Earth's bow shock: A comparison of upstream and downstream}. (a) Scaling exponent $\zeta$ of the pth-order structure function obtained by ESS fitting for the upstream and downstream Geotail datasets. The dashed line corresponds to K41 $q=q/3$ linear scaling. The error of the least-squares fitting is marked by the bars. (b) Quantification of amplitude-phase synchronization by kurtosis (top panel) and phase coherence index (bottom panel) for the upstream and downstream regions of the Earth's bow shock. The arrows a, b, and c correspond to scales $\tau =$1, 4, and 16 s, respectively.
         [reproduced from Ref. Koga et al. (2007)]}
        \label{fig:12}
    \end{figure}
     Koga et al. (\citeyear{koga2007intermittent}) performed a comparative study of the degree of intermittency (multifractality) and amplitude-phase synchronization of solar wind turbulence upstream and downstream of the Earth's bow shock. This study is based on the magnetic-field data of \textit{Geotail} from 18:00 UT 8 October to 04:00 UT 9 October 1995. In order to separate the data into upstream and downstream regions, the velocity and density data of \textit{Geotail} are used in conjunction with the bow shock model of Fairfield (\citeyear{fairfield1971average}). The computed scaling exponent $\zeta(p)$ in Fig. 12(a) shows that the degree of intermittency (multifractality) downstream of the Earth's bow shock is higher than upstream of the Earth's bow shock. This result is confirmed by kurtosis and phase coherence index in Fig. 12(b), which demonstrates that the degree of amplitude-phase synchronization in multiscale interactions downstream of the Earth's bow shock is higher than upstream of the Earth's bow shock. In addition, kurtosis and phase coherence index in Fig. 12(b) show that for large timescales the magnetic fluctuations are nearly Gaussian. For both upstream and downstream regions, the degree of amplitude-phase synchronization increases as the timescale decreases. This indicates the presence of coherent structures in the intermittent magnetic turbulence at both upstream and downstream of the Earth's bow shock.\\
\section{Theory}
    \subsection{Alfv\'en chaos, intermittency, and complexity}
    \begin{figure}[ht]
        \centering   
        \includegraphics[width=0.6\linewidth]{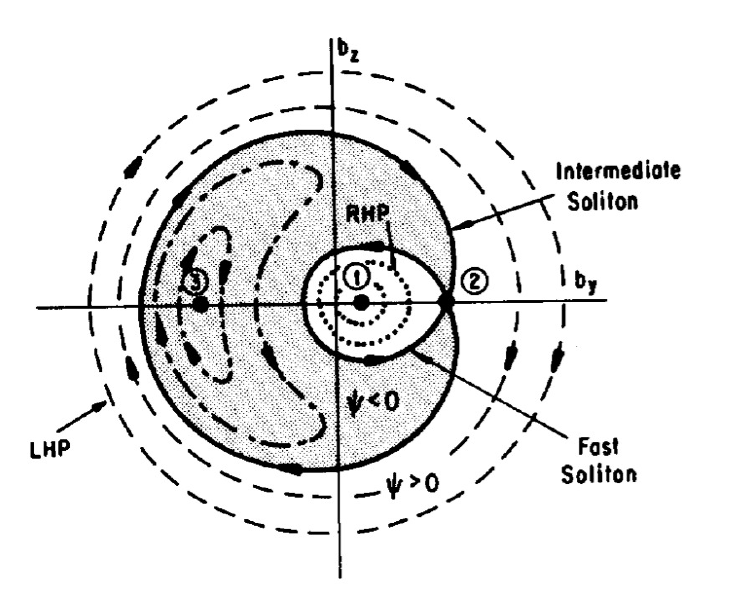}
        \caption{\textbf{Alfv\'en nonlinear waves and solitons.}The potential ($\psi$) level contours in the ($b_y$, $b_z$) phase space of stationary nonlinear Alfv\'en waves in the absence of a driver and dissipation.  The potential has a local maximum (labelled 1), a local minimum (labelled 3), and a saddle point (labelled 2). In this case, all solutions are regular. The solid lines denote the zero-energy soliton separatrices; the inner separatrix is the phase-space trajectory of the right-hand polarized (RHP, fast MHD mode) dark soliton when $\alpha$ is positive; the outer separatrix is the phase-space trajectory of the left-hand circularly polarized (LHP) bright soliton, which is in the intermediate MHD mode when $\alpha$ is positive. The soliton separatrices divide this phase space into three parts, so that there are three types of periodic nonlinear traveling waves. The dashed constant-energy contours denote purely left-handed waves; the dotted contours inside the dark soliton separatrix correspond to purely right-handed waves, and the dot-dashed contours inside the shaded region denote a wave that can have mixed right-left polarization, when its contour is near the soliton separatrices. [reproduced from Ref. Hada et al. (1990)]}
        \label{fig:13}
    \end{figure} 
\newpage
    \begin{figure}[ht]
        \centering
        \includegraphics[width=0.38\linewidth]{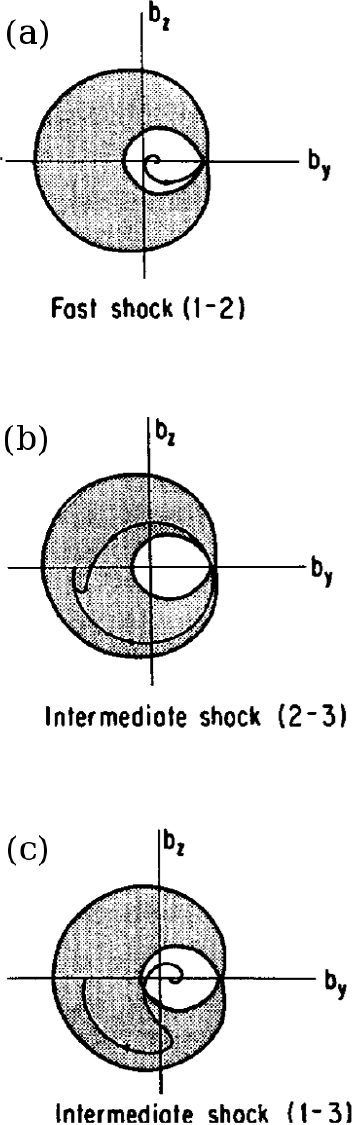}
        \caption{\textbf{Alfv\'en fast and intermediate shocks.}The potential level contours in the ($b_y$, $b_z$) phase space of stationary Alfv\'en shock waves in the presence of dissipation without a driver: (a) the 1-2 fast shock, (b) the pair of 2-3 intermediate shocks, (c) one member of the family of 1-3 intermediate shocks.  [reproduced from Ref. Hada et al. (1990)]}
        \label{fig:14}
    \end{figure}
   Nonlinear dynamics of Alfv\'en waves can be modelled by the derivative nonlinear Schr\"{o}dinger equation (DNLS) (Hada et al. \citeyear{hada1990chaos}; Chian et al. \citeyear{ chian1998alfven}; Borotto et al.  \citeyear{borotto2004alfven}; Rempel et al. \citeyear{rempel2006alfven})\\
   \begin{equation} \label{eq:1+Chian+ NPG 2007}
      \partial _{t}b+\alpha \partial _{x}\left( \abs{b}^{2}b\right) -i\left( \mu+i\eta \right) \partial _{x}^{2}b=S(b,x,t)\ ,
   \end{equation}
where the wave is propagating along an ambient magnetic field $B_0$ in the x-direction, $b=b_y+ib_z$ is the complex transverse wave magnetic field normalized to the constant ambient magnetic field, $\mu$ is the dispersive parameter, $\eta$ is a characteristic scale length, time $t$ is normalized to the inverse of the ion cyclotron frequency 
$\omega_{ci}=e B_{0}/m_{i} $, space $x$ is normalized to $c_{A}/\omega_{ci}$, $c_{A}=B_{0}/(\mu_{0}\rho_{0})^{1/2}$ is the Alfv\'en velocity, $c_S=(\gamma P_0/\rho_0)^{1/2}$ is the acoustic velocity, $\alpha=1/[4(1-\beta)]$, and $\beta=c^2_S /c^2_A$. The external forcing $S(b, x, t)=A \exp(i \,k\phi)$ is a monochromatic left-hand circularly polarized wave with a wave phase $\phi=x-V \,t$, where $V$ is a constant wave velocity, $A$ is the driver amplitude, and $k$ is the driver wave number. Equation (\ref{eq:1+Chian+ NPG 2007})   allows certain arbitrariness for choosing the signs of its various parameters (Ghosh  and Papadopoulos \citeyear{ghosh1987onset}; Chian et al. \citeyear{chian2007chaos}) and has been extensively used to study nonlinear MHD phenomena. We investigate the low-dimensional model of nonlinear Alfv\'en waves (Hada et al. \citeyear{hada1990chaos}, Chian et al. \citeyear{chian2007chaos}) by seeking traveling wave solutions of Eq. (\ref{eq:1+Chian+ NPG 2007})   with $b=b(\phi)$, whose first integral reduces to the following set of three nonlinearly coupled ordinary differential equations describing the transverse wave magnetic fields and the wave phase of nonlinear Alfv\'en waves\\
\begin{align} 
     \dot{b}_{y}-\nu \dot{b}_{z}&=\partial H/\partial b_{z}+a\cos \theta ,  \label{eq:3a+Hada+PF1990}\\
    {b}_z+\nu \dot{b}_y&=-\partial H/\partial b_y+a\sin \theta ,\label{eq:3b+Hada+PF1990}\\
    \dot{\theta}&=\Omega  ,\label{eq:3c+Hada+PF1990}
\end{align}
and 
\begin{equation}   \label{eq:3d+Hada+PF1990}
    H=({\bf {b}}^{2}-1)^{2}/4-(\lambda /2)({\bf {b}}-\hat{{\bf {y}}})^{2},
\end{equation}
where $b \rightarrow b/b_{0}$ (where $b_{0}$ is an integration constant), $\mathbf{b}=(b_{y}, b_{z})$, 
$w=b/b_{0}$, 
the normalized driver amplitude parameter $a=A/\alpha b^{2}\, k$, the normalized damping parameter
$ \nu=\eta/\mu$, the overdot denotes derivative with respect to the wave phase $\tau=\alpha b_{0}^2\phi/\mu$, $\theta=\Omega\phi$, $\Omega=\mu \,k/\alpha b_{0}^2$ , and $\lambda=-1+V/\alpha b^2_0$. We assume $\beta<1$ and $\nu$ is positive. Eqs. (\ref{eq:3a+Hada+PF1990})-(\ref{eq:3d+Hada+PF1990}) can be regarded as a model of  nonlinear driven-damped oscillator containing two control parameters, $a$ and $\nu$. 
\subsubsection{Alfv\'en nonlinear waves, solitons, and shocks }

In the absence of driving and dissipation, $a=\nu=0$, the number of dimensions of Eqs. (\ref{eq:3a+Hada+PF1990})-(\ref{eq:3d+Hada+PF1990})  reduces to two. In this case, the solutions of Eqs. (\ref{eq:3a+Hada+PF1990})-(\ref{eq:3d+Hada+PF1990}) are regular lying on constant potential energy (equipotential) surface contours of $h(w)= e$, where the value of $e$ is given by the initial conditions. 
Figure (13) shows the potential level contours for $\lambda=1/4$ and $\alpha$ positive in the ($b_y$, $b_z$) phase space appropriate to the problem. The potential has a local maximum (1), a local minimum (3), and a saddle point (2), such that $h(1)>h(2)>h(3)$, and $h(2)=0$. 
The shaded region shows where the potential $h$ is negative; the heavy solid lines threading the saddle point indicate where the potential is zero, and the potential approaches positive infinity as $\abs{w} \rightarrow \infty$. 
Each contour line of constant $e$ corresponds to one obliquely propagating elliptically polarized Alfv\'en wave, whose form and spatial period may be expressed in terms of elliptical integrals. The two infinite period orbits on the $h=0$ separatrices that connect to the saddle point have been called ``bright'' and ``dark'' solitons. 
The magnitude of the magnetic field increases in the bright soliton and decreases in the dark soliton. When $\alpha$ is positive, these solitons correspond to intermediate and fast MHD mode, respectively, as indicated in Fig. 13. The bright soliton is left-hand elliptically polarized, and the dark soliton is right-hand polarized when $\alpha$ is positive. 
The two soliton separatrices divide the phase space into three regions, hence there are three types of nonlinear periodic Alfv\'en waves. In the regions of positive potential, one of the waves is purely left-hand polarized (dashed $h=e$ contours beyond the bright soliton separatrix) and one is purely right-hand polarized (dotted contours inside the dark soliton separatrix). In the region of negative potential, the wave can have mixed right-left polarization (dot-dashed contours between the two soliton separatrices). 
Strictly speaking, this wave is mixed polarized only if its phase space contour lies close to the two soliton separatrices; if its contour closely encircles the potential minimum, it will be purely left-hand polarized. Since the average of the transverse field does not vanish in general, Alfv\'en nonlinear waves and solitons propagate obliquely to the ambient magnetic field.\\

In the presence of dissipation, the solutions can cross contours of constant $h$. The orbits then describe irreversible shock transitions connecting two out of three stationary points, corresponding, if $\alpha > 0$, to fast and intermediate MHD shocks, respectively. Figure 14 shows shock solutions computed numerically for $\alpha$ positive, $\lambda=1/4$, and $\nu=0.4$. 
The unique fast shock solution (Fig. 14(a)), which connects the stationary points 1 and 2, starts upstream at the potential high, and ends on the saddle point 2. Two types of nonunique intermediate shock solutions connect to the local minimum point 3. Figure 14(b) shows the pair of so-called intermediate shock solutions, which start upstream at the saddle point 2 and end downstream at the potential minimum point 3. Figure 14(c) shows one member of the one-parameter family of 1-3 intermediate shocks, which start at the potential high and end at the potential minimum.
\subsubsection{Hamiltonian chaos}
    \begin{figure}[ht]
        \centering
        \includegraphics[width=0.8\linewidth]{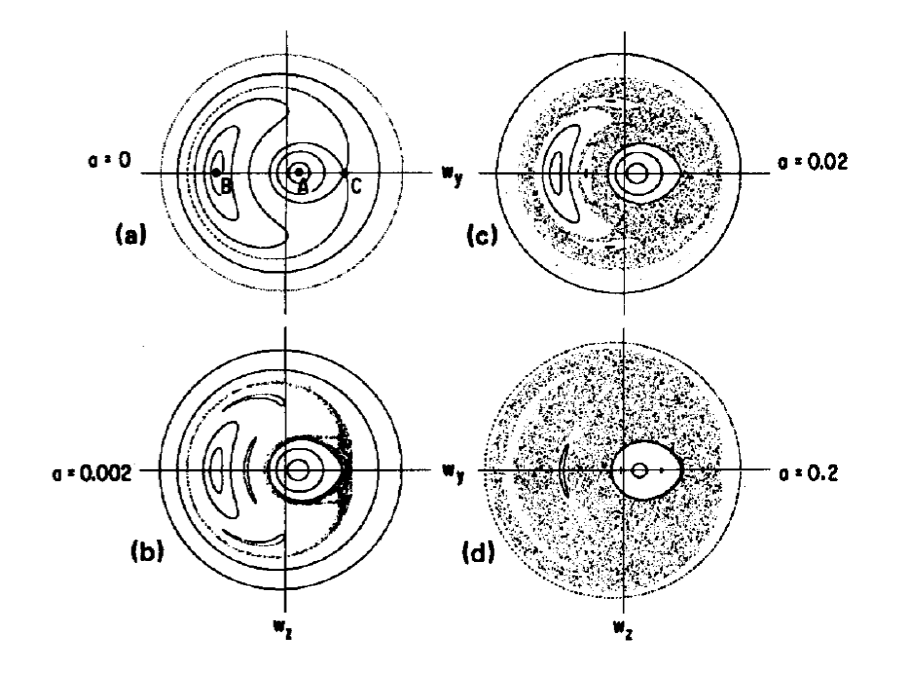}
        \caption{\textbf{Alfv\'en Hamiltonian chaos.}Poincar\'e  points in the ($w_y$, $w_z$) Poincar\'e plane of stationary driven Alfv\'en Hamiltonian system in the absence of dissipation: (a) $a = 0$, (b) $a =0.002$, (c) $a = 0.02$, (d) $a = 0.2$. The set of Poincar\'e points in (a) repeat the equicontour lines of Figure 13 with a local maximum (A), a local minimum (B), and a saddle point (c). In (a) there is no chaos, only periodic nonlinear waves and solitons as seen in Figure 13. Chaos appears even at the small driver amplitude in (b), evidenced by a chaotic orbit near the soliton separatrices. At the same time, one of the dark wave trajectory separates into three islands which yields another group of chaotic orbits near the separatrices surrounding these islands.  As the driver amplitudes increases, the area including chaotic orbits expands, as seen in (c) and (d).  [reproduced from Ref. Hada et al. (1990)]}
        \label{fig:15}
    \end{figure}

In the presence of driving and absence of dissipation, the number of dimensions of Eqs. (\ref{eq:3a+Hada+PF1990})-(\ref{eq:3d+Hada+PF1990}) becomes three, which is the minimum degree of freedom required for chaos to occur.  Hence, in this case the solutions of Eqs. (\ref{eq:3a+Hada+PF1990})-(\ref{eq:3d+Hada+PF1990}) can be chaotic. The phenomenon of Alfv\'en Hamiltonian chaos is illustrated in Fig. 15, which shows the solutions of Fig. 14 in the Poincar\'e map by projecting the solution on the ($w_y$, $w_z$) phase space once for each period $2\pi/\Omega$ of the wave phase $\tau$, i.e,,  defining a Poincar\'e plane as\\ 
$$P : [b_y (\tau ), b_z(\tau )] \rightarrow [b_y (\tau + T ), b_z(\tau + T )],$$
where $T =2\pi/\Omega$ is the driver period. Figure 15(a) shows the regular solutions in the absence of driving ($a=0$), when the entire set of Poincar\'e points originating from a given initial point remains on the potential contour containing that initial point, since the contour is the cross section of a torus whose surface contains the solution orbit. For a small-amplitude driver ($a=0.002$), Fig. 15(b) shows one of the sets of Poincar\'e points near the ``bright'' soliton separatrix starts to scatter in a limited region of the phase phase, indicating the onset of Hamiltonian chaos. It is well known that perturbation  of a generic Hamiltonian system first yields chaotic motion in a layer surrounding a separatrix, because the perturbed manifolds of trajectories coming into and out of the saddle point (hyperbolic point) start to cross each other and making infinitely many intersections (homoclinic points). As a result, the trajectories near the saddle point become enormously complicated and the corresponding chaotic Alfv\'en waveforms exhibit sudden, unpredictable jumps in the wave phase as well as sense of polarization. In addition, Fig. 15(b) shows that chaotic orbits can result from higher-order resonances, e.g., one of the orbits near the local minimum of the potential breaks up into three resonance islands, indicating that there exist three hyperbolic points between these islands. Orbits starting sufficiently close to any of these hyperbolic points will be chaotic. Note, however, most of the area in the phase space of Fig. 15(b) remain regular. The chaotic regions are separated from the regular regions by KAM tori  (Lichtenberg and Lieberman \citeyear{lichtenberg1983regular}). Figures 15(c)-(d) show that as the driver amplitude $a$ increases further, the chaotic region expands in the phase space which is an evidence of strong chaos. At $a=0.2$, Fig. 15(d) shows that most of the phase space is chaotic except a few small islands, reaching a state of global stochasticity. Note that the region of right-hand polarization inside the dark soliton separatrix remains regular in Figs. 15(d), even when almost all the remainder phase space is chaotic. This is because the left-hand driver can induce stochasticity only in the regions where the polarizations of the unperturbed periodic waves are also left-hand polarized, or have a mixed left- and right-hand polarization.

\subsubsection{Dissipative chaos: Crisis-induced intermittency}
    \begin{figure}[ht]
        \centering
        \includegraphics[width=0.5\linewidth]{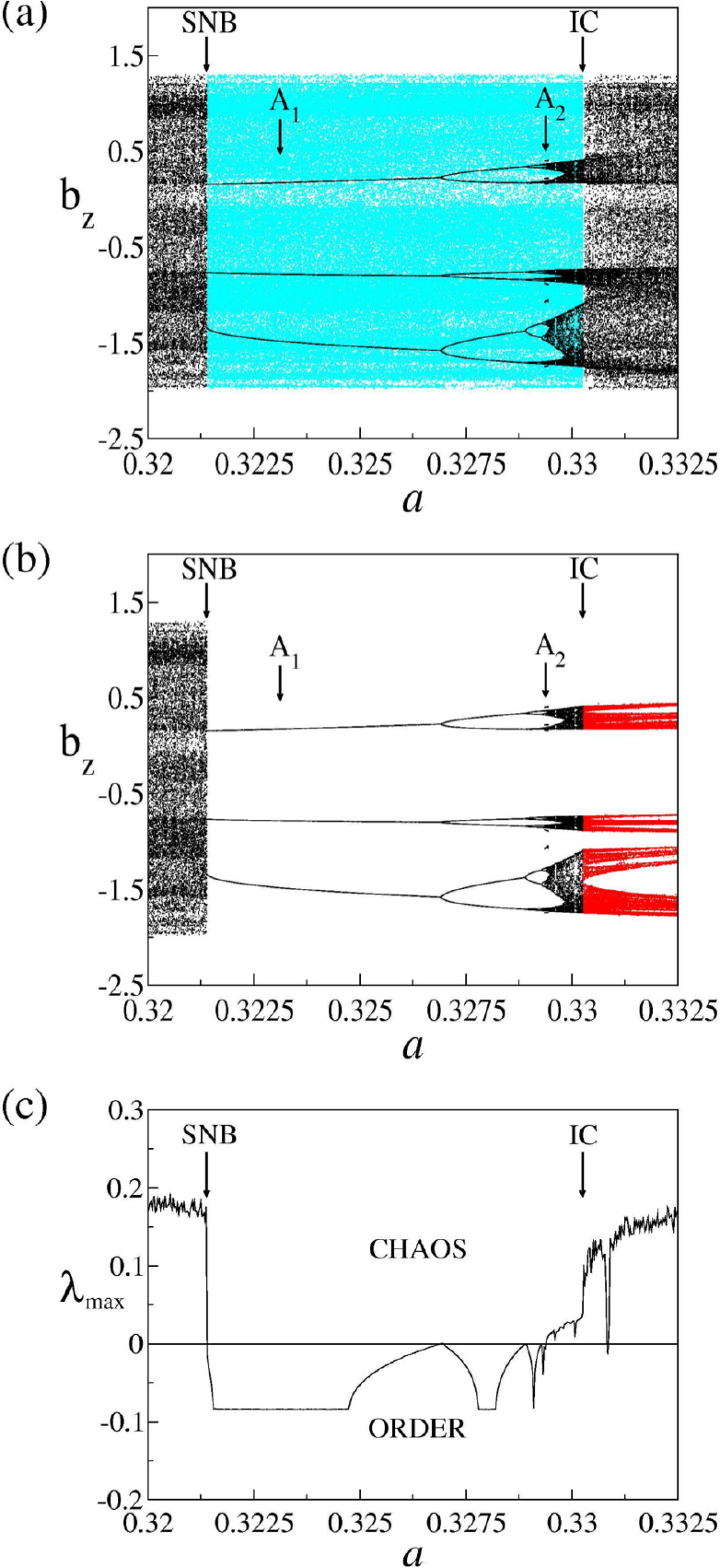}
        \caption{\textbf{Alfv\'en bifurcation diagram and maximum Lyapunov exponent: Periodic attractor, chaotic attractor, and chaotic saddle.} Limit point diagram and maximum Lyapunov exponent: period-3 periodic window. (a) Limit point diagram, $b_z$ as a function of the driver amplitude $a$, for attractors $A_1$ and $A_2$, superimposed by the surrounding chaotic saddle (blue); (b) the same as (a), showing the conversion of the pre-crisis banded chaotic attractor (black) into the post-crisis banded chaotic saddle (red); (c) maximum Lyapunov exponent, $\lambda_{max}$ as a function of $a$, for the attractor $A_1$. SNB denotes saddle-node bifurcation, and IC denotes interior crisis.  [reproduced from Ref. Chian et al. (2007)]
        }
        \label{fig:16}
    \end{figure}
    \begin{figure}[ht]
        \centering
        \includegraphics[width=0.45\linewidth]{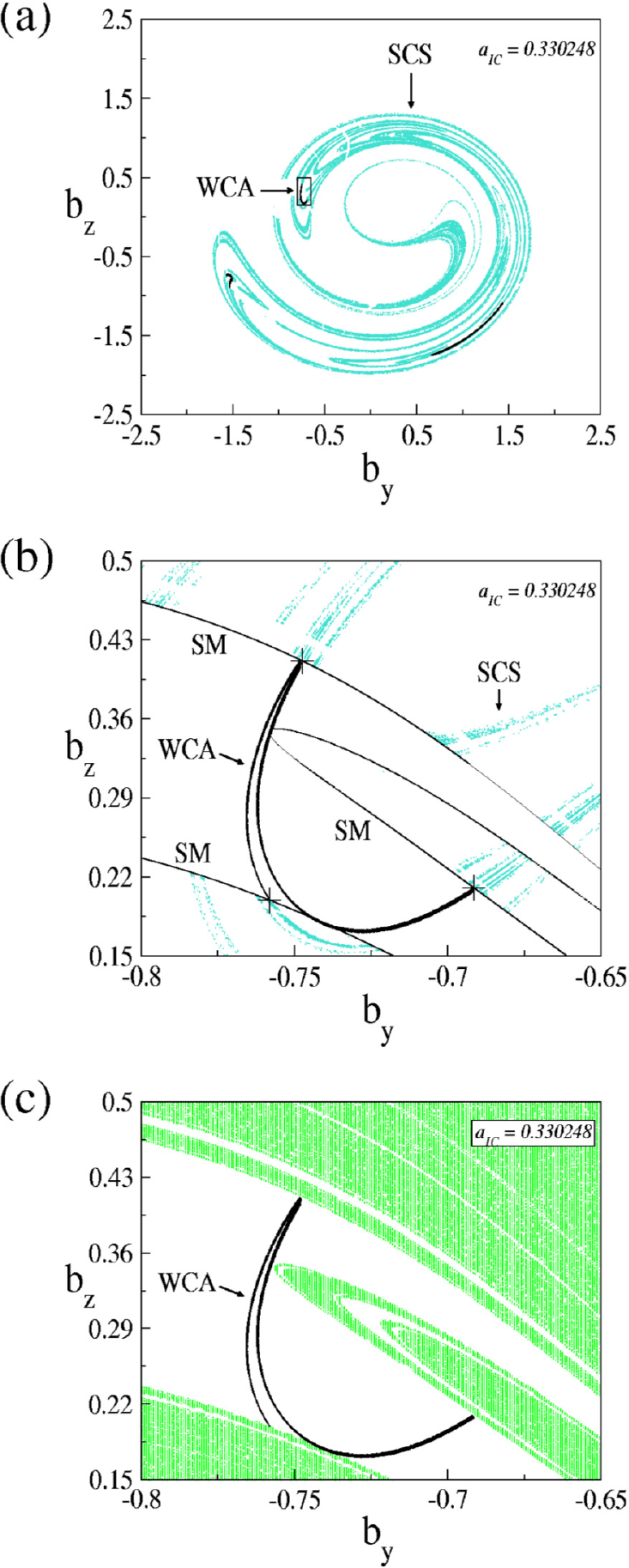}
        \caption{\textbf{Alfv\'en interior crisis: Chaotic attractor-chaotic saddle collision at interior crisis for} $\mathbf{a_{IC}=0.330248}$. (a) Pre-IC surrounding chaotic saddle (SCS, blue) and weak (banded) chaotic attractor (WCA, black); (b) a zoom of (a) showing the collision of the weak chaotic attractor (WCA) with the mediating period-9 unstable periodic orbit (cross), its stable manifold (SM) and the surrounding chaotic saddle (SCS); (c) same as (b) showing the collision of the weak chaotic attractor (WCA) with the stable manifold (green) of the surrounding chaotic saddle. [reproduced from Ref. Chian et al. (2007)]}
        \label{fig:17}
    \end{figure}
    \begin{figure}[ht]
        \centering
        \includegraphics[width=0.5\linewidth]{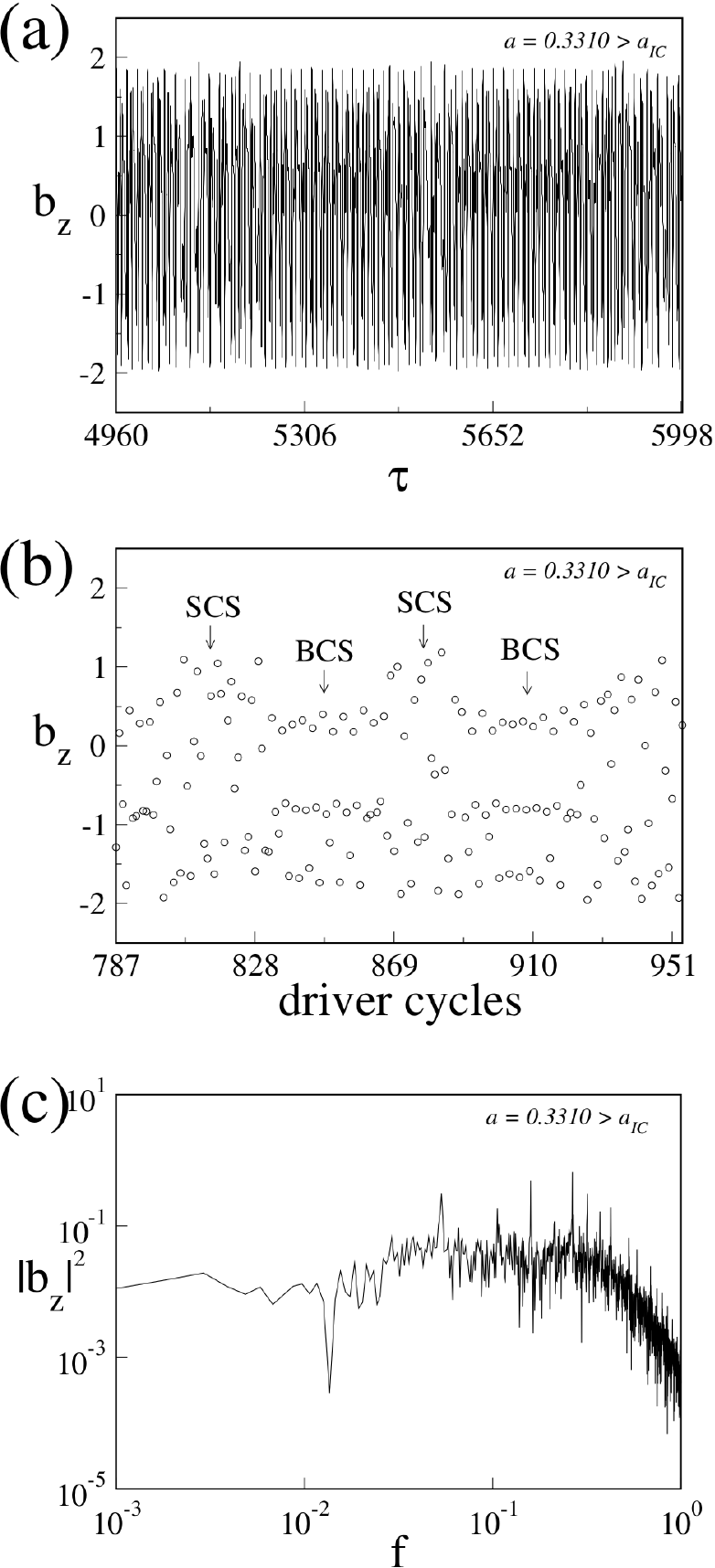}
        \caption{\textbf{Alfv\'en crisis-induced intermittency.} Alfv\'en crisis-induced intermittency. (a) Times series $b_z$ as a function of $\tau$ for $a=0.3310$, (b) same time series as (a) plotted as a function of driver cycles, (c) power spectrum of (a). SCS denotes surrounding chaotic saddle and BCS denotes banded chaotic saddle. [reproduced from Ref. Chian et al. (2007)]}
        \label{fig:18}
    \end{figure}
    In the presence of driving and dissipation, the solutions of Eqs. (\ref{eq:3a+Hada+PF1990})-(\ref{eq:3d+Hada+PF1990})  admit chaotic attractors as well as chaotic saddles, both consisted of an infinite set of unstable periodic orbits. A bifurcation diagram, providing an overview of the system dynamics and its sensitive dependence on small variations in a system parameter, can be constructed from the numerical solutions of Eqs. (\ref{eq:3a+Hada+PF1990})-(\ref{eq:3d+Hada+PF1990}) by varying the driver amplitude $a$ while keeping other systems parameters fixed ($\nu=0.02, \Omega=-1, \lambda=1/4, \mu=1/2$). Figure 16(a) gives an example of bifurcation diagram of nonlinear Alfv\'en waves, showing a  periodic window where two attractors $A_1$ and $A_2$ are found. For a given $a$, Fig. 16(a) plots the asymptotic values (black dots) of the Poincar\'e points of $b_z$. This periodic window begins with a saddle-node bifurcation (SNB), where a pair of period-3 stable and unstable periodic orbits appear. The period-3 stable periodic orbit ($A_1$) undergoes a cascade of period-doubling bifurcations as $a$ increases and turns eventually into a banded chaotic attractor with three bands. This periodic window ends with an interior crisis (IC). Moreover, two chaotic saddles (blue and red) are shown in Figures 16(a)-(b) by plotting a straddle trajectory close to a chaotic saddle computed from the PIM triple algorithm (Nusse and York \citeyear{ nusse1989procedure}; Rempel and Chian \citeyear{rempel2004alfven}). The blue region inside the periodic window denotes the surrounding chaotic saddle (SCS) which extends to the chaotic regions outside the periodic window, to the left of SNB and to the right of IC, where it becomes a subset of the chaotic attractor. This chaotic saddle is called a surrounding chaotic saddle because it ``surrounds'' the phase space occupied by the attractors within the periodic window as well as the phase space occupied by the banded chaotic saddle after the interior crisis.     After the interior crisis, the banded chaotic attractor loses its stability and is converted into a banded chaotic saddle (red) as seen in Fig. 16(b). The maximum Lyapunov exponent of the attractor $A_1$ is shown in Fig. 16(c).\\
    
    Interior crisis is a global bifurcation that involves the conversion of a weak (banded) chaotic attractor into a strong chaotic attractor (Grebogi et al. \citeyear{ grebogi1983crises}; Borotto et al.  \citeyear{borotto2004alfven}), characterized by an abrupt increase of the maximum Lyapunov exponent as seen in Fig. 16(c). Figure 17(a) shows the coexistence of the weak chaotic attractor (WCA, black) and the surrounding chaotic saddles (SCS, blue) just prior to IC in the Poincar\'e plane. The interior crisis is driven by the collision of a weak chaotic attractor with a surrounding chaotic saddle (Figs. 17(b)-(c)), mediated by a period-9 unstable periodic orbit created by a saddle-node bifurcation responsible for the appearance of the attractor $A_2$. The phenomenon of interior crisis depicted in Fig. 17(b) is similar to the phenomenon of edge of chaos in the laminar-turbulent transition (Chian et al. \citeyear{chian2013edge}), where the mediating unstable periodic orbit (cross) plays the role of the edge state, and its stable manifold (SM) plays the role of the edge of chaos. As the result of chaotic attractor-chaotic saddle collision, a strong chaotic attractor appears after the onset of interior crisis as the result of the coupling between weak and strong chaotic saddles as well as a set of coupling unstable periodic orbits newly created by the phenomenon of explosion at the gap regions (shown in Fig. 17(c)) of chaotic saddles after the onset of interior crisis.\\
    
    Alfv\'en intermittency can be generated by a saddle-node bifurcation responsible for type-I Pomeau-Manneville intermittency (Chian et al. \citeyear{chian1998alfven}, \citeyear{chian2006chaotic}), or by an interior crisis responsible for crisis-induced intermittency. Figures 18 (a)-(b) show the time series of Alfv\'en crisis-induced intermittency after the interior crisis described in Fig. 17, whereby episodic regime switching between laminar and bursty phases of the magnetic field fluctuations are observed. The laminar (bursty) phase corresponds to the trajectory traversing near the weak (strong) chaotic saddle, respectively. This intermittent regime switching between small-amplitude fluctuations and spiky bursts as well as the power-law behaviour of the power spectrum (Fig. 18(c)) reproduce the temporal dynamics of Alfv\'en intermittent turbulence observed in the solar wind discussed in Section 3.\\
\subsubsection{Noise-induced intermittency}
    \begin{figure}[ht]
        \centering
        \includegraphics[width=\linewidth]{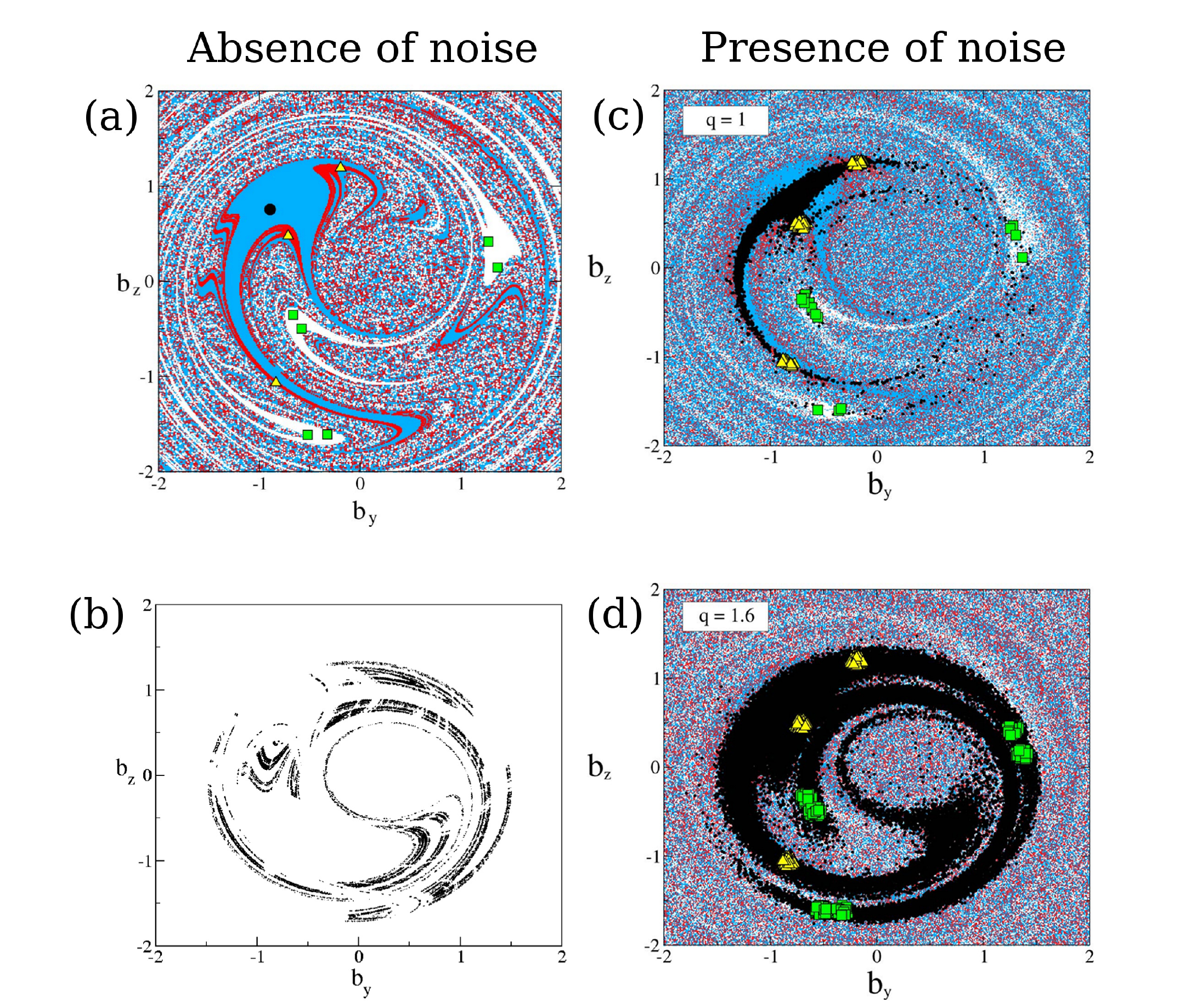}
        \caption{\textbf{Alfv\'en chaotic saddle, basins of attraction, and noise.} Basins of attraction and Poincar\'e points of periodic attractors (a) and chaotic saddle (b) in the absence of noise; basins of attraction and Poincar\'e points in the presence of noise ((c) and (d)). (a) Periodic attractors $A_1$ (circle), $A_2$ (triangles) and $A_4$ (squares) and their basins of attraction in blue ($A_1$), red ($A_2$) and white ($A_4$), at $\nu = 0.01746$; (b) Chaotic saddle on the boundary separating the basins of attraction at $\nu = 0.01746$. Basins of attraction and Poincar\'e points of the intermittent noisy trajectories at $\nu = 0.01746$ for Gaussian noise $q = 1$ (c); and for non-Gaussian noise $q = 1.6$ (d). In both cases, $\sigma_{q} = 0.064$. Triangles are plotted whenever the orbit is in the vicinity of attractor $A_2$; squares refer to the vicinity of attractor $A_4$; black circles represent points in the vicinity of attractor $A_1$ or the surrounding chaotic saddle. [reproduced from Ref. Rempel et al. (2008)]}
        \label{fig:19}
    \end{figure}
    \begin{figure}[ht]
        \centering
        \includegraphics[width=0.45\linewidth]{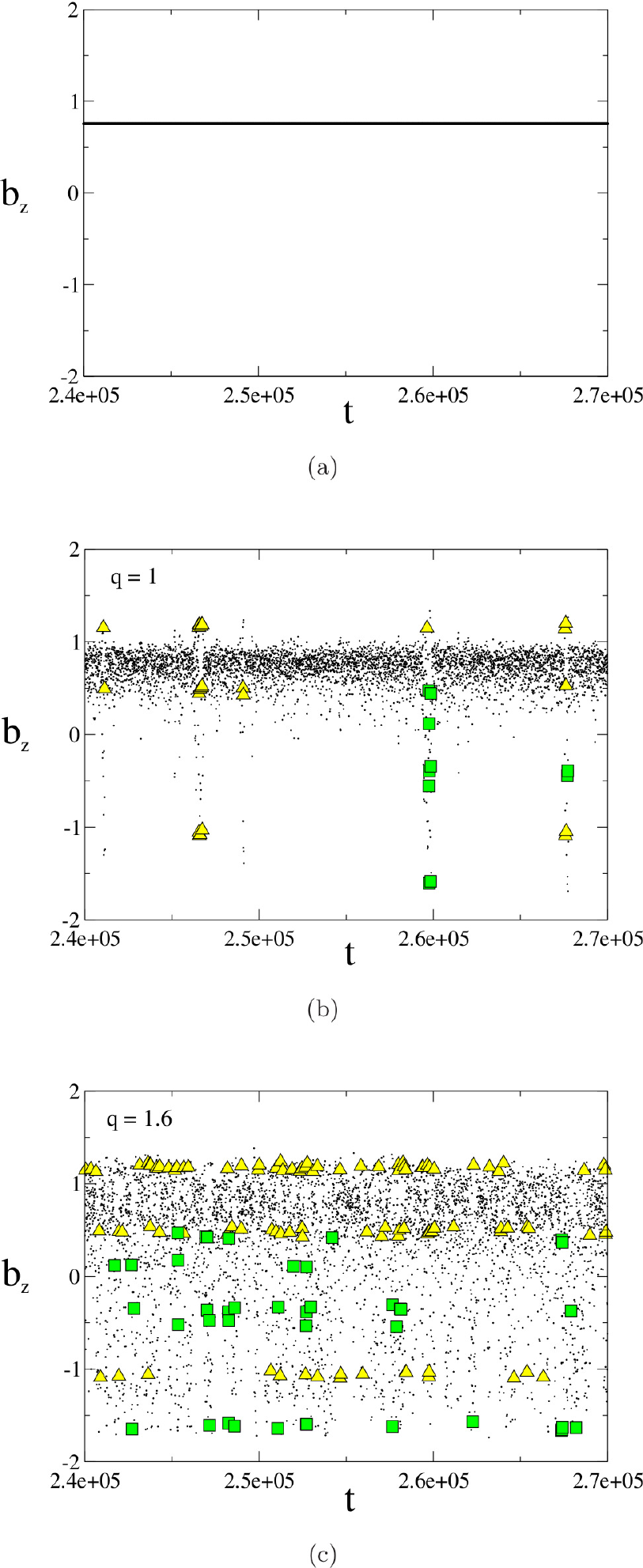}
        \caption{\textbf{Alfv\'en noise-induced intermittency.} (a) Time-$2\pi$ time series of $b_z$ in terms of the driver cycles for period-1 attractor $A_1$ at $\nu = 0.01746$. (b) Intermittency induced by Gaussian noise, for $\sigma_{q}= 0.064$ and $q = 1$. Triangles are plotted whenever the orbit is in the vicinity of attractor $A_2$ and squares refer to the vicinity of $A_4$. Black circles represent points in the vicinity of $A_1$ or the surrounding chaotic saddle. (c) Intermittency induced by non-Gaussian noise for $\sigma_{q}= 0.064$ and $q = 1.6$.
        [reproduced from Ref. Rempel et al. (2008)]}
        \label{fig:20}
    \end{figure}
    Up to now we have only considered deterministic Alfv\'en systems. In reality, Alfv\'en intermittent turbulence in space and laboratory plasmas is an admixture of deterministic and stochastic systems. The complex dynamics of nonlinear Alfv\'en waves described by the driven-dissipative DNLS was investigated by Rempel et al. (\citeyear{rempel2006alfven}) and Rempel et al. (\citeyear{rempel2008alfven}) by introducing additive Gaussian and non-Gaussian noise, respectively, in the governing equations Eqs.  (\ref{eq:3a+Hada+PF1990})-(\ref{eq:3d+Hada+PF1990}). Multistability is a common property of complex systems. As mentioned in Section 2.2, it can be an obstacle for prediction, since the asymptotic state may depend crucially on the initial condition. The bifurcation diagram in Fig. 16(a) shows that there is coexistence of two attractors. Rempel et al. (\citeyear{rempel2008alfven}) studied a region of the bifurcation diagram of Eqs. (\ref{eq:3a+Hada+PF1990})-(\ref{eq:3d+Hada+PF1990}) where five attractors are found. 
The complexity of this multistable region is best depicted in Fig. 19(a), which shows the Poincar\'e map of three out five coexisting periodic attractors $A_1$ (circle, period-1), $A_2$ (triangles, period-3), and $A_4$ (squares, period-6) and their basins of attraction at $\nu = 0.01746$. The basin of attraction is the set of initial conditions in the ($b_y$, $b_z$) phase plane which converge to a given attractor.
The blue region in Fig. 19(a) is the basin of $A_1$, red represents the basin of $A_2$ and white the basin of $A_4$. 
The basin boundaries display a complex structure where coherent regions are observed around the periodic attractors and incoherent regions permeate the surrounding phase space, where the three basins seem to mingle. This complex structure is scale invariant, a typical property of fractal sets. Rempel et al. (\citeyear{rempel2008alfven}) numerically found a chaotic saddle embedded in the fractal basin boundary.  The chaotic saddle in Fig. 19(b) detected by the sprinkler method (Hsu et al. \citeyear{hsu1988strange}) plays an important role in attractor hopping and Alfv\'en noise-induced intermittency. 
In order to model the stochastic dynamical system to interpret the observation of space plasma turbulence, Rempel et al. (\citeyear{rempel2008alfven}) introduced an external stochastic source in Eqs. (\ref{eq:3a+Hada+PF1990})-(\ref{eq:3d+Hada+PF1990}) by adding a non-Gaussian noise based on the Tsallis nonextensive statistical mechanics (Tsallis \citeyear{tsallis1988possible}).
Figure 20 illustrates the effect of noise on the Poincar\'e time series of $b_z$ at  $\nu = 0.01746$. Figure 20(a) shows the noise-free time-$2\pi$ time series of the period-1 attractor $A_1$ in terms of the driver cycles. 
As shown by Rempel et al. (\citeyear{rempel2006alfven}), in the presence of noise $A_1$ resembles a chaotic attractor, stretching along directions for which the attraction is weakest. 
For a small-amplitude noise, the perturbed trajectories stay confined to the vicinity of $A_1$. If the noise level is strong enough to stretch $A_1$ beyond its basin boundary, the Alfv\'en wave ``escapes''  from the basin, wanders for a certain amount of time in the complex boundary region before settling to a different attractor, leading to the occurrence of attractor hopping (Arecchi and Lisi \citeyear{arecchi1982hopping}).  
For a Gaussian noise with zero mean and standard deviation $\sigma$, the trajectories can escape from the basin of attraction of $A_1$ when $\sigma \geqslant 0.064$. 
The noise will then trigger attractor hopping until the trajectory eventually is reinjected into the basin of $A_1$. 
This process repeats intermittently, generating the noise-induced intermittency (Gwinn and Westervelt \citeyear{ gwinn1985intermittent}), shown in Fig. 20(b). Most of the time the value of the $b_z$ component of the Alfv\'en magnetic field oscillates around 0.78, in the vicinity of $A_1$. 
There are several intermittent ``bursts'' to lower values of $b_z$, indicating an excursion of the trajectory through a different region of the ($b_y$, $b_z$) phase space. 
The triangles and squares indicate when the trajectory is in the vicinity of attractors $A_2$ and $A_4$, respectively. We consider a vicinity defined as the disk with radius equal to $\sigma$ around the fixed points of the periodic attractors in the Poincar\'e map. 
Each attractor has an associated time scale, i.e., a mean escape time for a trajectory to leave its neighbourhood. Figure 20(c) shows the intermittent time series obtained with a non-Gaussian noise, with a q-standard deviation  $\sigma_{q} = 0.064$ and an arbitrarily chosen nonextensivity parameter $q = 1.6$. 
The occurrence of intermittent bursts is greatly increased due to the fat tails of the non-Gaussian PDF. Note that in every burst in Figs. 20(b)-20(c) there are some points which are not in the vicinity of either $A_2$ or $A_4$. 
Those points represent the time the trajectory spends around the complex basin boundary region associated with the chaotic saddle, before converging to the vicinity of an attractor. This dynamics is shown in Poincar\'e maps in Figs. 19(c)-(d), which plot the noisy basins of attraction and the Poincar\'e points corresponding to the time series of Figs. 20(b)-(c). 
For a Gaussian noise (Fig. 19(c)), most points concentrate in a stretched region around $A_1$ and the scattered points represent the intermittent bursts. 
For a non-Gaussian noise (Fig. 19(d)), the stochastic component seems to dominate the system dynamics. However, a comparison between Figs. 19(c)-(d) and Fig. 19(b) reveals that in each burst the trajectory visits the neighbourhood of the chaotic saddle. 
This occurs because, although the chaotic saddle is not attracting, it possesses a stable manifold, which is a zero measure set in the phase plane whose points display trajectories which converge to the chaotic saddle (Nusse and Yorke \citeyear{nusse1989procedure}).\\

\subsection{Chaos in parametric wave-wave interaction}

    \begin{figure}[h]
        \centering
        \includegraphics[width=0.6\linewidth]{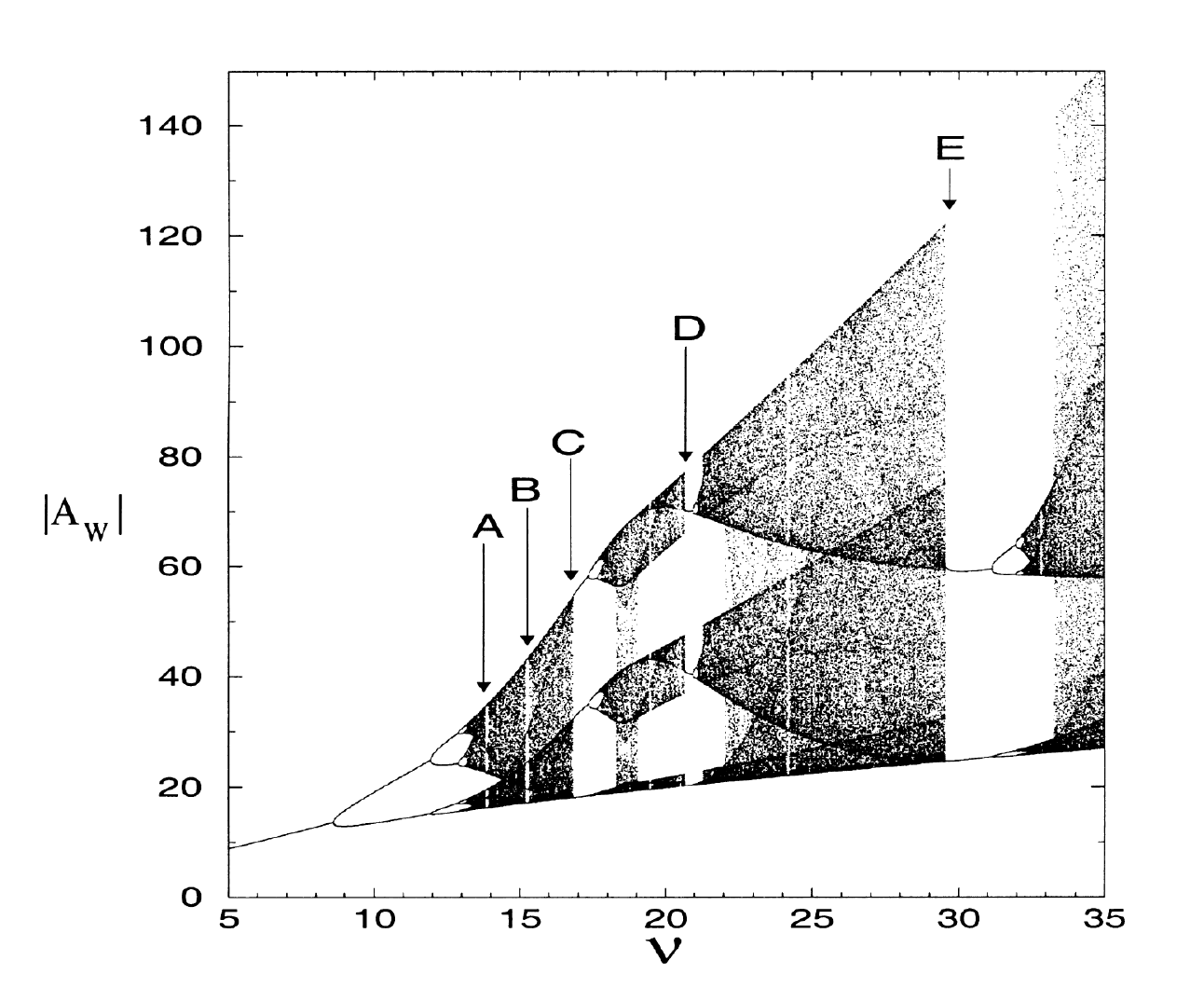}
        \caption{\textbf{Bifurcation diagram: Parametric wave-wave interaction.} Bifurcation diagram of $\abs{ A_{W}}$ as a function of $\nu$ for $\delta = 2$. Type-I Pomeau-Manneville intermittency occurs at $\nu \sim 13.81$ (A), 15.21 (B), 16.82 (C), 20.64 (D) and 29.56 (E).
        [reproduced from Ref. Chian et al. (2000)]}
        \label{fig:21}
    \end{figure}%
    \begin{figure}[ht]
        \centering
        \includegraphics[width=0.6\linewidth]{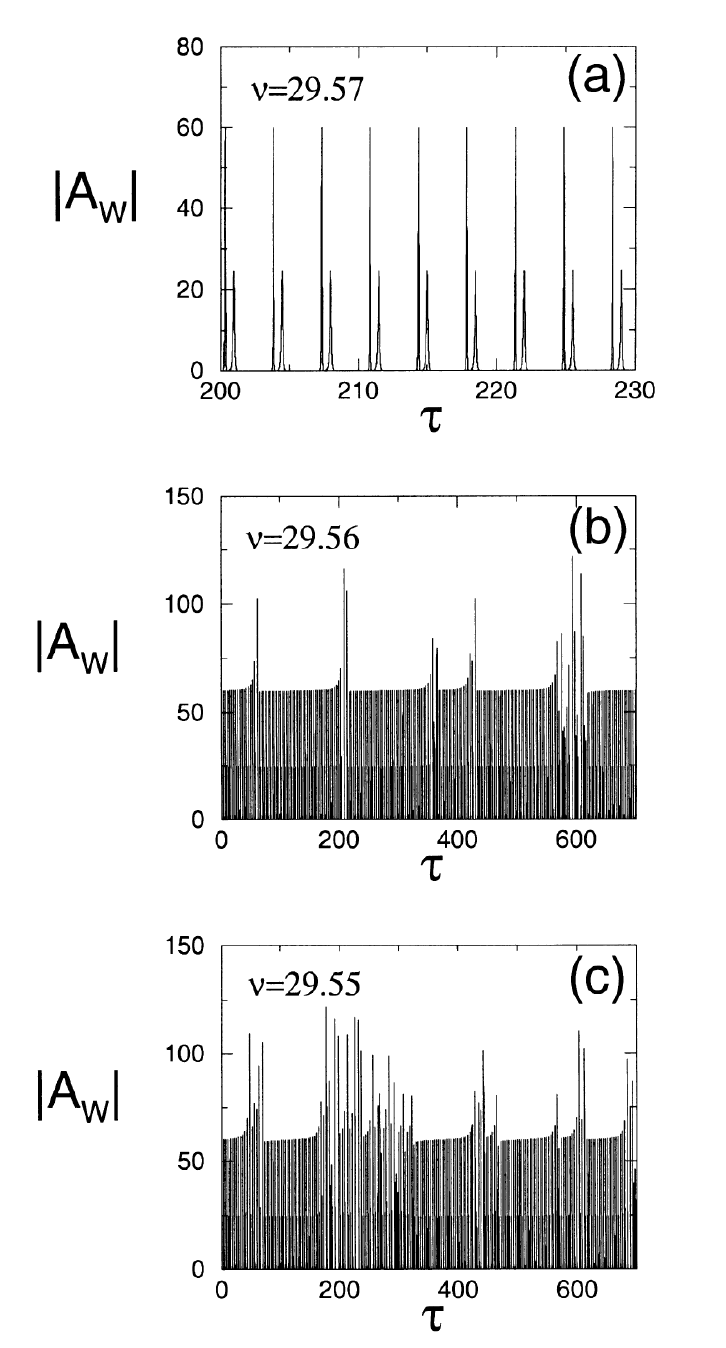}
        \caption{\textbf{Type-I Pomeau-Manneville intermittency: Parametric wave-wave interaction.} Time series of $\abs{A_{W}}$ of the type-I Pomeau-Manneville intermittency for $\nu =$  (a) 29.57, (b) 29.56 and (c) 29.55. [reproduced from Ref. Chian et al. (2000)]}
        \label{fig:22}
    \end{figure}
    As an example of parametric wave-wave interaction, we consider three-wave parametric decay of a Langmuir wave (L) into a whistler wave (W) and an Alfv\'en wave (A), all traveling along the ambient magnetic field $B=B_o \hat{z}$, which meet the following phase-matching conditions 
    \begin{equation}  \label{eq:1+pss2000}
         \omega_{L} \approx \omega_{W} + \omega_{A}, \hspace{1cm} 
          \mathbf{ k}_{L} = \mathbf{ k}_{W} + \mathbf{k}_{A},
    \end{equation}
\noindent
where a frequency mismatch and a perfect wave vector match are assumed. In addition to the wave frequency and wavevector matching conditions of Eq. (\ref{eq:1+pss2000}), the wave triplet must also satisfy the conservation of wave helicity. Since the electromagnetic whistler wave is right-hand circularly polarized, the Alfv\'en wave is left-hand circularly polarized (i.e., shear Alfv\'en mode). If we look for  traveling wave solutions, the nonlinear system of three coupled wave equations can be written in the normalized form \\

\begin{align}  
     \dot{A}_{L} &= \nu_{L} A_L + A_W A_A, \label{eq:21+pss2000}  \\
     \dot{A}_{W} &= \nu_{W} A_W - A_L A_A^*, \label{eq:22+pss2000}\\ 
     \dot{A}_{A} &= i \delta A_A +\nu_{A} A_A - A_L A_W^*,  \label{eq:23+pss2000}
\end{align}
\noindent
 where $A_L$,  $A_W$, and $A_A$ are the wave amplitude of Langmuir, whistler, and Alfv\'en wave, respectively; the dot denotes differentiation with respect to the phase variable $\tau=k(z-Vt)$, $V$ and $k$ are arbitrary wave velocity and wave vector, respectively; $\nu_L$ is the linear growth parameter representing unstable Langmuir wave driven by an electron beam-plasma instability, $ \nu_W (\nu_A)$ is the damping parameter of whistler (Alfv\'en) wave and we assume $\nu_W=\nu_A\equiv -\nu < 0$; $\delta$ is the frequency mismatch parameter.\\
 
A bifurcation diagram for the solutions of Eqs. (\ref{eq:21+pss2000})-(\ref{eq:23+pss2000}) is shown in Fig. 21 by varying the control parameter $\nu$ (wave growth) and keeping other parameters fixed, which contains a wealth of dynamical behaviours including divergence, fixed point, limit cycle (periodic attractor), and chaotic attractor (Wersinger  et al. \citeyear{wersinger1980bifurcation}; Meunier et al. \citeyear{meunier1982intermittency}). 
Five periodic windows are indicated in Fig. 21, where (A, B, C, D, E) denotes the beginning of each periodic window characterized by a saddle-node bifurcation, responsible for the route to chaos when a periodic attractor loses its stability and is converted into a chaotic attractor. 
After a transition to chaos via saddle-node bifurcation, the system retains the memory of the periodic attractor, hence the chaotic system exhibits an episodic regime switching between periods of laminar and bursty fluctuations. 
The laminar regime corresponds to the trajectory at the vicinity of the periodic attractor prior to the saddle-node bifurcation, whereas the bursty regime corresponds to the surrounding chaotic saddle (similar to Fig. 16).
 Figures 22(b)-(c) show time series of the type-I Pomeau-Manneville intermittency resulting from the saddle-node bifurcation that occurs at the control parameter $\nu=29.56$ (E) in Fig. 21:  Fig. 22(a) shows the periodic attractor with period-2 prior to the saddle-node bifurcation; Fig. 22(b) shows the chaotic attractor just after the saddle-node bifurcation; and Fig. 22(c)  shows the chaotic attractor further away from the saddle-node bifurcation. It is worth mentioning that the results discussed in this section apply to other tree-wave parametric decay interactions if they are governed by the same set of nonlinearly coupled quadratic equations  Eqs. (\ref{eq:21+pss2000})-(\ref{eq:23+pss2000}). \\

\subsection{Chaos in modulational wave-wave interaction}
    \begin{figure}[ht]
        \centering
        \includegraphics[width=\linewidth]{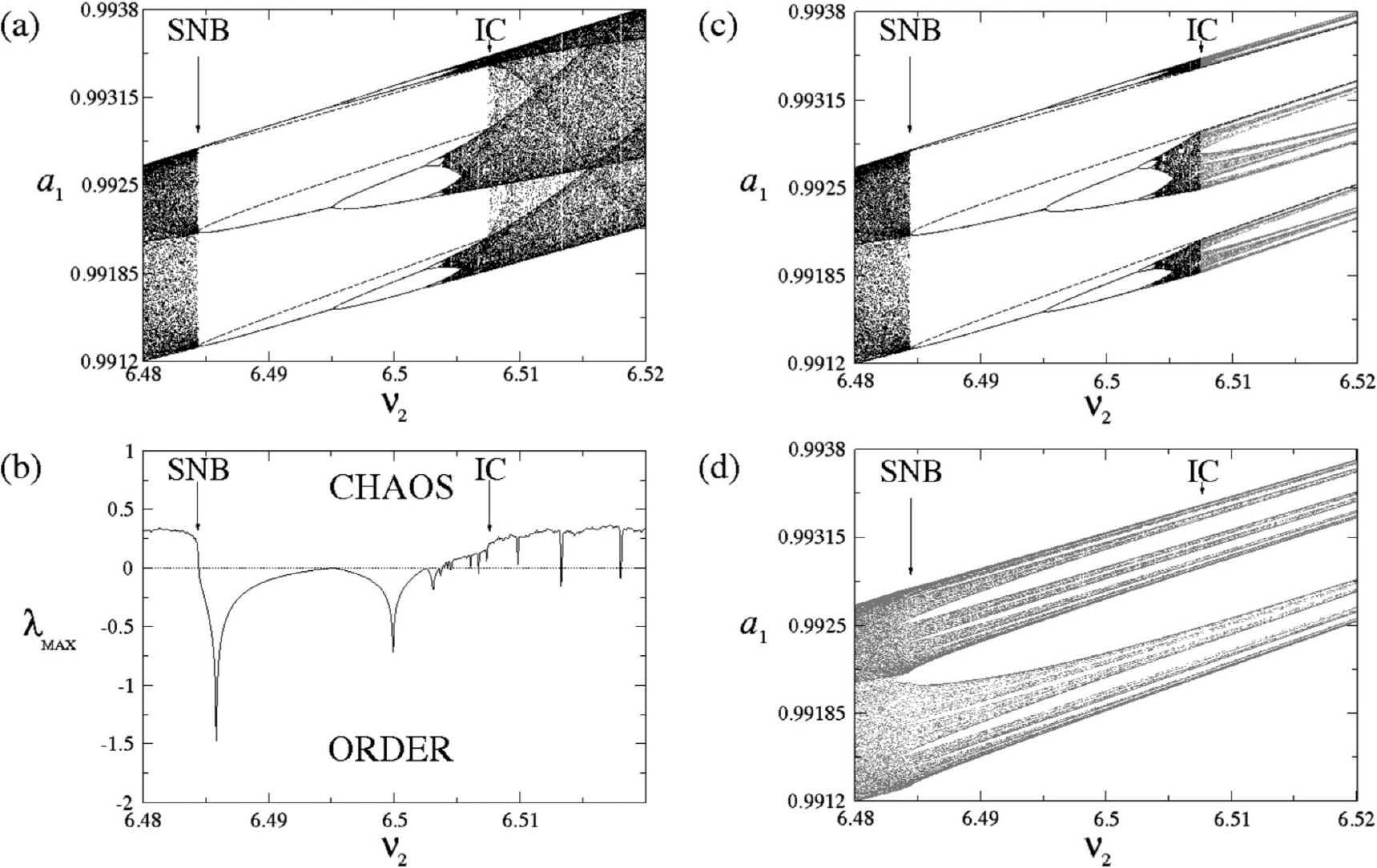}
        \caption{\textbf{Bifurcation diagram and maximum Lyapunov exponent: Modulational wave-wave interaction.}(a) Bifurcation diagram for $a_{1}$ as a function of $\nu_{2}$. The dashed lines represent the evolution of the period-3 unstable periodic orbit, SNB denotes a saddle-node bifurcation and IC denotes an interior crisis. (b) Maximum non-zero Lyapunov exponent $\lambda_{MAX}$ as a function of $\nu_2$. (c) Bifurcation diagram showing the conversion of the weak chaotic attractor (black) to a banded chaotic saddle (gray) and its evolution after crisis. (d) Bifurcation diagram showing the evolution of the surrounding chaotic saddle.
        [reproduced from Ref. Miranda et al. (2012)]}
        \label{fig:23}
    \end{figure}
    \begin{figure}[ht]
        \centering
        \includegraphics[width=0.6\linewidth]{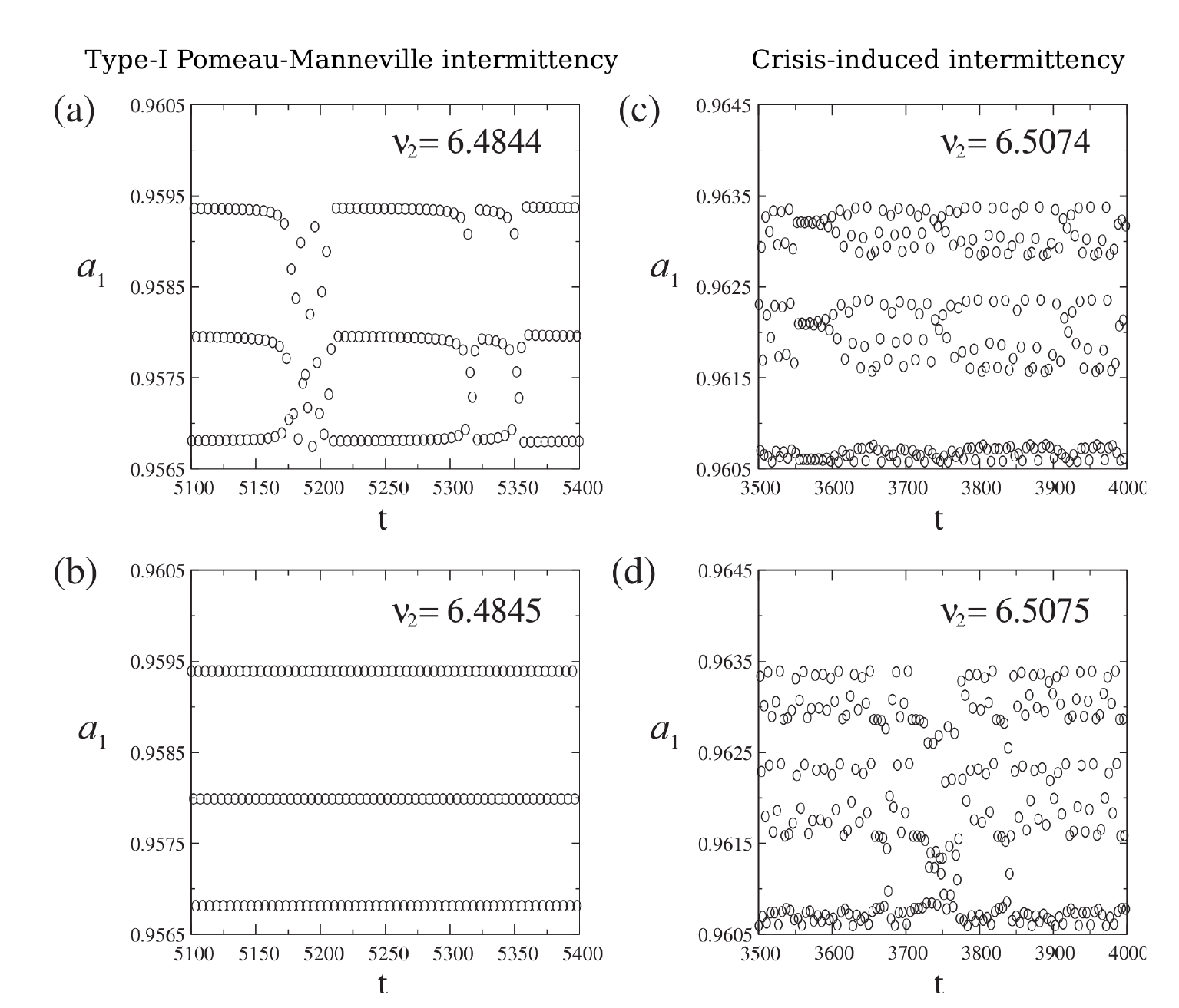}
        \caption{\textbf{Type-I Pomeau-Manneville intermittency and crisis-induced intermittency: Modulational wave-wave interaction.} (a) Time series of the Poincar\'e points of $a_1$ at $\nu_2 = 6.4844$, showing the characteristics of  type-I Pomeau-Manneville intermittency. (b) Time series of the Poincar\'e points of $a_{1}$ at $\nu_{2} = 6.4845 > \nu_{SNB}$, showing the dynamics of a periodic attractor with period-3. (c) Time series of pre-crisis weak chaos represented by the Poincar\'e points of  $a_{1}$ at $\nu_{2} = 6.5074  < \sim \nu_{IC}$. (d) Time series of  $a_1$ at $\nu_{2} = 6.5075 >\sim \nu_{IC}$ represented by the Poincar\'e points, showing a crisis-induced intermittency alternating episodically between periods of weak and strong chaos.
        [reproduced from Ref. Miranda et al. (2012)]}
        \label{fig:24}
    \end{figure}
    Modulational wave-wave interaction (Miranda et al. \citeyear{miranda2012chaotic}) can be described by the nonlinear Schr\"odinger equation governing the evolution of the slow-varying envelope of the wave field
    
        \begin{equation}   \label{eq:7+Miranda+PoP2012}
         i(\partial_t E + \hat{\nu} E + \partial_z^2 E + \abs{E}^{2} E) = 0
    \end{equation}
    
\noindent
where $\hat{\nu}$ denotes the linear growth/damping rate of \textit{E}. We look for traveling wave solutions for the three-wave truncation of Eq. (\ref{eq:7+Miranda+PoP2012}) that satisfy the resonant condition
\begin{equation}
     2k_0 = k_1 + k_2 ,
\end{equation}
where the subscript 0 denotes a linearly growing pump wave, 1 and 2 denote linearly damped Stokes and anti-Stokes daughter waves, respectively. By assuming linear dispersion relations for the waves, we obtain the following nonlinear system of coupled wave equations 
\begin{align}   
   \dot{a}_0&= \nu_0 a_0+2 a_0 a_1 a_2 \sin\phi ,   \label{eq:12+Miranda+PoP2012}\\    
    \dot{a}_1&= -\nu_1 a_1- a^2_0 a_2 \sin\phi , \label{eq:13+Miranda+PoP2012}\\  
    \dot{a}_2&= -\nu_2 a_2- a^2_0 a_1 \sin\phi , \label{eq:14+Miranda+PoP2012}\\
\end{align}
\begin{equation}     \label{eq:15+Miranda+PoP2012}
    \dot{\phi}=-2\delta + a_1^2+a_2^2-2 a_0^2 + \left[4 a_1 a_2 - a_0^2\left(\dfrac{a_2}{a_1}+\dfrac{a_1}{a_2}\right) \right]\cos \phi,
\end{equation}
where the dot denotes derivative with respect to the normalized time, $\phi(t) = 2\psi_0 - \psi_1 - \psi_2 – 2{\delta}t$, $\nu_0=-\hat{\nu}(k_0), \hat{\nu}_{1,2}=\hat{\nu}(k_{1,2})$. Equations (\ref{eq:12+Miranda+PoP2012})-(\ref{eq:15+Miranda+PoP2012}) can describe  modulational wave-wave interaction in Alfv\'en turbulence (Ghosh and Papadopoulos \citeyear{ghosh1987onset}) and Langmuir turbulence (Miranda et al. \citeyear{miranda2012chaotic}). 
In terms of quantum mechanics, Eqs. (\ref{eq:12+Miranda+PoP2012})-(\ref{eq:15+Miranda+PoP2012}) describe the nonlinear temporal evolution of a 4-quanta system wherein a pair of pump quanta interact with a pair of Stokes and anti-Stokes daughter quanta. 
We analyze a period-3 periodic window of the bifurcation diagram of Fig. 23(a). Similar to Fig. 16(a) this periodic window is created by a saddle-node bifurcation at $\nu_2 = \nu_{SNB} \sim 6.4845$, where an order-to-chaos transition occurs and a pair of period-3 stable and unstable orbits are created. 
The period-3 stable periodic orbit  undergoes a cascade of period-doubling bifurcation leading to onset of chaos when the maximum Lyapunov exponent becomes positive as shown in Fig. 23(b). The three-banded chaotic attractor undergoes an interior crisis at $\nu_2 \sim 6.5025$ when it collides with the surrounding chaotic saddle given by Fig. 23(d) and the mediating period-3 unstable periodic orbit created at the saddle-node bifurcation as seen in Fig. 23(a), and turns into a strong chaotic attractor. 
Figure 23(b) shows that there is a sudden increase in the value of the maximum Lypaunov exponent at interior crisis. After the interior crisis,  the weak chaotic attractor loses stability and turns into a banded chaotic saddle shown in Fig. 23(c), confined to the same banded region as the pre-crisis weak chaotic attractor.  
Furthermore, there is a surrounding chaotic saddle for the entire bifurcation diagram as shown in Fig. 23(d). The chaotic saddles are obtained using the PIM-triple algorithm (Nusse and Yorke \citeyear{nusse1989procedure}). Both banded and surrounding chaotic saddles in Figs. 23(c)-(d) contain empty regions called ``gaps'' that widen as the control parameter increases. 
After the interior crisis, the gap regions are densely filled with newly created coupling unstable periodic orbits that have components of Poincar\'e points located in both banded and surrounding chaotic saddles. 
This process of gap filling is an example of a bifurcation phenomenon called ``explosion''. \\

Type-I Pomeau-Manneville intermittency appears to the left of saddle-node bifurcation and crisis-induced intermittency appears to the right of interior crisis, as illustrated by the Poincar\'e time series in Fig. 24. Both types of intermittency display episodic switching between laminar regime of small-amplitude fluctuations and bursty regime of large-amplitude fluctuations. 
The laminar regime corresponds to the trajectory of wave solution passing by the vicinity of the period-3 unstable periodic orbit (banded chaotic saddle) for type-I Pomeau-Manneville intermittency (crisis-induced intermittency). 
The bursty regime corresponds to the trajectory passing by the vicinity of the surrounding chaotic saddle for both types of intermittency. 
The gap filling is responsible for the crisis-induced intermittency shown in Figs. 24(c)-(d), because the trajectory can escape from the gap region of a chaotic saddle to the gap region of the other chaotic saddle via the unstable manifolds of a coupling unstable periodic orbit. It is worth pointing out that the results discussed in this section apply to other four-wave modulational interactions if they are governed by the same set of nonlinearly coupled cubic equations (Eqs. (\ref{eq:12+Miranda+PoP2012})-(\ref{eq:15+Miranda+PoP2012})).
\section{Discussion and conclusion}
  Space, astrophysical, and laboratory plasmas are governed by ubiquitous and universal stochastic and nonlinear dynamical processes. Many chaotic, complex, and intermittent turbulence phenomena observed in space plasmas are also observable in astrophysical and laboratory plasmas. For example, flares and coronal mass ejections have been detected in stars and the Galactic Center of Milky Way. There is a growing interest in stellar coronal mass ejections and flares due to their potential impact on stellar evolution and the exoplanet habitability. Flares in our nearest stellar neighbour Proxima Centauri (Davenport et al. \citeyear{Davenport2016most}) and superflares in solar-type stars (Notsu et al. \citeyear{notsu2019kepler})  have been observed. Evidence of an eruptive filament from a superflare in a young solar-type star (Namekata et al. \citeyear{namekata2021probable})  and coronal mass ejections through coronal dimming of a cool star (Veronig et al. \citeyear{veronig2021indications})  have been reported. The nonlinear methods discussed in Section 3 for investigating space observations of  magnetic reconnection and intermittent turbulence in coronal mass ejections in solar atmosphere, solar wind, and at the Earth’s bow shock,  can be applied to study extra-solar coronal mass ejections and their impact on the exoplanets (Chian et al. \citeyear{chian2010planetary}). Very Large Array (VLA) centimeter-wavelength radio imaging of Galactic Center in the region near the supermassive black hole Sagittarius A* (Sgr A*) identified a variety of filamentary and coherent structures (LaRosa et al. \citeyear{larosa2000wide}), e.g., thread, mouse, snake, and tornado, which indicates that the plasma environment surrounding Sgr A* is in a turbulent state dominated by intermittent magnetic structures, likely related to coherent structures such as magnetic flux ropes and current sheets prevalent in space plasma turbulence discussed in Section 3. X-ray and infrared observations of Sgr A* display highly-variable intermittent behaviour,  with X-ray flares rising above a quiescent thermal background about once per day (Boyce et al. \citeyear{boyce2019simultaneous}). Atacama Large Millimeter/submillimeter Array (ALMA) observations of light curve of Galactic Center in 2019 June, when Sgr A* underwent strong flaring activity in the near-infrared brightening by up to a factor of 100 compared to quiescent values, suggest that the brightest near-infrared flares of Sgr A* are likely caused by magnetic reconnection (Murchikova eet al. \citeyear{murchikova2021second}). General-relativistic MHD simulations (Ripperda et al. \citeyear{ripperda2022black}) confirmed that plasmoid-mediated magnetic reconnection can power flares originating from the inner magnetosphere of accreting black holes; magnetic reconnection near the event horizon produces sufficiently energetic plasma to explain flares from accreting black holes such as the TeV emission observed from M87, the giant elliptical galaxy whose nucleus contains the first supermassive black hole ever directly imaged.\\
  
  Jets and current sheets have been observed in the region of magnetic reconnection in laser-produced plasma experiments using the OMEGA facility (Rosenberg et al. \citeyear{rosenberg2015slowing}). Spiky electric and magnetic fields resulting from magnetic reconnection of multiple magnetic flux ropes were seen in the experiments performed in the large plasma device (LAPD); a Jensen-Shannon complexity-entropy analysis discussed in Section 3 shows that the spiky nonlinear structures are chaotic (Gekelman et al. \citeyear{gekelman2019spiky}). Choi et al. (\citeyear{choi2021effects}) investigated the effects of edge plasma turbulence on the nonlinear evolution of magnetic island in the Korea Superconducting Tokamak Advanced Research (KSTAR) experiment. The uncontrolled magnetic island is a serious problem in tokamak devices since it often leads to plasma disruption. The turbulence and shear flow developed around a magnetic island produce complex transport behaviour at the boundary layers of the magnetic island. Moreover, turbulence spreading into the magnetic island can enhance the turbulence level at magnetic reconnection site which can either retard or facilitate magnetic reconnection. The observations of Choi et al. (\citeyear{choi2021effects}) render support for the theoretical model of Alfv\'en Hamiltonian chaos (Hada et al. \citeyear{hada1990chaos}) depicted in Fig. 15, which shows that the onset of chaos (turbulence) appears at the magnetic soliton separatrices (corresponding to the magnetic island boundary layers). In particular, it elucidates the symbiotic relationship between magnetic reconnection and intermittent turbulence at the boundary layers of a magnetic flux rope (e.g., CME) and the interface region of multiple magnetic flux ropes (CME-CME merger) discussed in Section 3. This confirms the ubiquitousness of turbulent magnetic reconnection (Lazarian et al. \citeyear{lazarian20203d}) that plays a key role in energizing space, astrophysical, and laboratory plasmas.\\
  
  In addition to the laboratory experiment on magnetic reconnection reported by Gekelman et al. (\citeyear{gekelman2019spiky}), a number of other plasma laboratory experiments have also obtained evidence of chaos based on the Jensen-Shannon complexity-entropy analysis discussed in Section 3. Maggs and Morales (\citeyear{maggs2013permutation}) performed a basic experiment on electron heat transport in a magnetized afterglow plasma which established the chaotic nature of the underlying dynamics that causes anomalous transport. Maggs et al. (\citeyear{maggs2015chaotic}) showed that plasma density fluctuations in low confinement (L-mode) plasmas in the DIII-D tokamak are chaotic. In an experiment on intermittent magnetic turbulence in Swarthmore Spheromak that exhibits a Kolmogorov $5/3$ scaling, Schafner et al. (\citeyear{schaffner2016possible}) found that the dissipation mechanism in plasma is chaotic. In the turbulent plasma of a reversed-field pinch RELAX, Onchi et al. (\citeyear{onchi2017permutation}) showed that the complexity-entropy of soft-X ray and UV emissions and magnetic fluctuations depends on the conditions of plasma confinement; in the high-density regime it is close to the stochastic region, whereas in the low-density regime it approaches the chaotic region. Time series analysis by Zhu et al. (\citeyear{zhu2017chaotic}) of the Alcator C-Mod tokamak revealed that the turbulent edge density fluctuations are chaotic which is supported by observation of exponential power spectra associated with Lorentzian-shaped pulses in the time series. The aforementioned papers show that the chaos theory discussed in Section 4 can be applied to interpret chaotic phenomena observed in laboratory plasmas.\\
  
  In contrast to the observation of chaos in laboratory plasma turbulence, several studies have concluded that fluctuations in space plasma turbulence are stochastic in nature as shown in the complexity-entropy analysis by Miranda et al. (\citeyear{miranda2021complexity}) of magnetic fluctuations in interplanetary rope-rope magnetic reconnection discussed in Section 3. Weck et al. (\citeyear{weck2015permutation})  and Oliver, Engelbrecht, and Strauss (\citeyear{olivier2019permutation}) showed that solar wind magnetic fluctuations for both fast and slow streams measured by \textit{Wind} and \textit{ACE} at 1 AU are stochastic with complexity-entropy values close to pure white noise and more random than even classical Brownian motion; the fast solar stream signal exhibits slightly more entropy and less complexity than the slow solar stream signal. The complexity-entropy study by Osmane et al. (\citeyear{osmane2019jensen}) showed that the \textit{AL} geomagnetic auroral index, that provides an estimate of the maximum westward auroral electrojet intensity, is indistinguishable from stochastic processes from time scales ranging from a few minutes to 10 h. A comprehensive statistical analysis of  solar wind magnetic structures by Weygand and Kivelson (\citeyear{weygand2019jensen}) that includes interplanetary coronal mass ejections, co-rotating interaction regions, and turbulent magnetic fluctuation intervals found that the turbulent intervals observed by \textit{Helios}, \textit{Wind}, and \textit{Ulysses} lie within the stochastic region of the complexity-entropy maps and that their complexity decreases while their normalized entropy increases with distance from the Sun. Good et al. (\citeyear{good2020radial}) performed a complexity-entropy analysis of the magnetic field time series in shock-sheath and upstream solar wind of an ICME event at \textit{MESSENGER} at 0.47 AU and subsequently by \textit{STEREO-B} at 1.08 AU while the two spacecraft were radially aligned. Their results show a trend of reducing complexity with radial distance, and an increased complexity in the shock-sheath intervals relative to the upstream solar wind, thus confirming the fractal dimension analysis of Mu\~noz et al. (\citeyear{munoz2018evolution}) which indicates that the shock-sheaths in two ICME events are more complex than upstream solar wind and the magnetic flux rope of ICME driver, and is in agreement with the study of the Earth's bow shock by Koga et al. (\citeyear{koga2007intermittent}) discussed in Section 3 which shows that the degree of amplitude-phase synchronization and multifractality downstream (magnetosheath) is higher than upstream. This finding is also consistent with an increased complexity in the stream interaction regions relative to the unperturbed solar wind found by Weygand and Kivelson (\citeyear{weygand2019jensen}). A greater complexity in ICME shock-sheaths compared to the upstream solar wind is in-line with our understanding of shock-sheath plasmas being dominated by a large number of coherent magnetic structures such as small-scale current sheets and magnetic flux ropes. Note that space plasma turbulence can behave as a dynamical system that is very sensitive to small variations of system parameters such as noise, hence the noise-induced intermittency discussed in Section 4 can readily appear in space plasmas even for a low level of noise. It is plausible that the fluctuations of space plasma turbulence are a combination of stochastic and chaotic dynamics.\\
  
  The dynamical systems approach to turbulence, such as chaotic saddles and Lagrangian coherent structures, provides powerful tools to unravel the complex dynamics of fluids and plasmas (Bohr et al. \citeyear{bohr1998dynamical}; Chian et al. \citeyear{chian2003dynamical}; Lai and T\'el \citeyear{lai2011transient}; Haller \citeyear{haller2015lagrangian}). Unstable and stable manifolds, discussed in Section 4, of fluid and plasma particles constitute distinguished material lines or surfaces that act as transport barriers in turbulence. These distinguished lines are the hyperbolic Lagrangian coherent structures that attract or repel the neighbouring material, both retarding and facilitating transport fluxes through chaotic mixing (Haller \citeyear{haller2015lagrangian}). Hence, they are responsible for organizing and mediating the transport and interaction of matter and energy in turbulent fluid and plasma flows. In particular, the attracting and repelling hyperbolic Lagrangian coherent structures act as transport barriers that enable the formation of the elliptic Lagrangian coherent structures (e.g., vortices, magnetic islands, magnetic flux ropes). Silva et al. (\citeyear{da2002escape}) showed that chaotic saddles can account for the appearance of chaos at ergodic magnetic limiters in the plasma-wall interaction region of tokamaks, which creates channels for fast escape of chaotic magnetic field lines. Padberg et al. (\citeyear{padberg2007lagrangian}) demonstrated that the heteroclinic tangles formed by the intersections of attracting and repelling Lagrangian coherent structures may explain the turbulent transport in magnetized fusion plasmas. Pegoraro et al. (\citeyear{pegoraro2019coherent}) discussed the application of Lagrangian coherent structures in the evolution of magnetic reconnection and showed that in the linear phase two independent magnetic island chains are formed at their resonant surfaces; as the reconnection instability grows the dynamics of the magnetic configuration becomes nonlinear, leading to the expansion of these chains; when the magnetic islands start to interact the regions where magnetic field lines are chaotic spread, similar to the Alfv\'en Hamiltonian chaos scenario illustrated in Fig. 15 and the KSTAR observation of Choi et al. (\citeyear{choi2021effects}). Di Giannatale et al. (\citeyear{di2021prediction}) confirmed that Lagrangian coherent structures are useful for identifying the hidden paths governing the chaotic motion of magnetic field lines and predicting the location of temperature gradients in reversed field pinch experiments; inside the chaotic region, the motion of magnetic field lines is far from stochastic. \\
  
  D\'emoulin et al. (\citeyear{demoulin1996three}) showed that quasi-separatrix layers, related to stable manifolds (for finite-time) discussed in Section 4, are thin layers of magnetic reconnection sites where the gradient of the mapping of magnetic field lines from one part of a boundary to another is very large; the relative thickness of the quasi-separatrix layers depends on the maximum twist of the magnetic flux tubes; the shape of the quasi-separatrix layers is typical of the two ribbons observed in two-ribbon solar flares, confirming that the accompanying prominence eruption involves the reconnection of twisted magnetic structures. Rempel et al. (\citeyear{rempel2011lagrangian}) and Rempel et al. (\citeyear{rempel2017objective}) showed that the dynamo turbulence in an MHD simulation is dominated by chaotic entanglement of attracting and repelling hyperbolic Lagrangian coherent structures, detected by computing the backward and forward finite-time Lyapunov exponent, respectively; elliptical Lagrangian coherent structures (i.e., vortices), detected by the technique of Lagrangian averaged vorticity deviation (Rempel et al. \citeyear{rempel2017objective}), are surrounded by separatrices given with the intersections of attracting and repelling Lagrangian coherent structures. Yeates et al. (\citeyear{yeates2012lagrangian}) showed how the build-up of magnetic gradients in the solar corona may be inferred directly from a 12 h \textit{Hinode} dataset of the horizontal photospheric velocity in a plage region of AR 10930, by computing the repelling Lagrangian coherent structure which corresponds to a network of quasi-separatrix layers in the magnetic field. Chian et al. (\citeyear{chian2014detection}) established the correspondence of the network of high magnetic flux concentration to the attracting Lagrangian coherent structures in the photospheric velocity based on the same \textit{Hinode} dataset used by Yeates et al. (\citeyear{yeates2012lagrangian}) and numerical simulations. Silva el at. (\citeyear{silava2018objective}) showed that the technique of Lagrangian averaged vorticity deviation (Rempel et al. \citeyear{rempel2017objective}) can detect vortices accurately in a 15 min time interval of \textit{Hinode} images of the quiet-Sun photosphere. Chian et al. (\citeyear{chian2019supergranular}) used a 7 hr \textit{Hinode} dataset of the quiet-Sun photosphere to show that the Lagrangian centers and boundaries of solar supergranular cells are given by the local maximum of the forward and backward finite-time Lyapunov exponent, respectively. The attracting Lagrangian coherent structures expose the location of the sinks of photospheric flows at supergranular junctions, whereas the repelling Lagrangian coherent structures interconnect the Lagrangian centers of neighbouring supergranular cells. Lagrangian transport barriers are found within a supergranular cell and from one magnetoconvective cell to other cells, which play a key role in the dynamics of internetwork and network magnetic elements. Such barriers favour the formation of persistent (recurrent) vortices in the complex mixed-polarity regions of supergranular junctions, at the footpoints of magnetic flux tubes/ropes that can lead to flares and coronal mass ejections via magnetic reconnection (Attie et al. \citeyear{attie2016relationship}; Chian et al. \citeyear{chian2020lagrangian}). The magnetic field distribution in the quiet Sun is determined by the combined action of attracting and repelling Lagrangian coherent structures and vortices. Chian et al. (\citeyear{chian2020lagrangian}) used a 22 h \textit{Hinode} dataset of the quiet-Sun photosphere to  report observational evidence of Lagrangian chaotic saddles in plasmas. A set of 29 persistent objective vortices with lifetimes varying from 28.5 to 298.3 min are detected by computing the Lagrangian averaged vorticity deviation. The unstable manifold of the Lagrangian chaotic saddles computed for $\approx$11 h exhibits twisted folding motions indicative of recurring vortices in a complex magnetic mixed-polarity region. In particular, it was shown that the persistent objective vortices are formed in the gap regions of Lagrangian chaotic saddles at supergranular junctions.\\
  
  In conclusion, space plasmas provide a natural laboratory for understanding the fundamental characteristics of chaos, complexity, and intermittent turbulence in nature. Knowledge on the formation and evolution of coherent structures such as soliton, vortex, magnetic island, magnetic flux rope, current sheet, and their interaction with turbulence via magnetic reconnection in space plasmas can help us to probe similar stochastic and nonlinear dynamical processes in astrophysical and laboratory plasmas.
  
\bmhead{Acknowledgments}
R.A.M. acknowledges financial support from FAP DF under award number 180/2020 and DPI/DPG/UnB from Brazil. ELR acknowledges financial support from Brazilian agencies CAPES and CNPq (Grant 306920/2020-4).
\bibliography{main}
\bmhead{Affiliations}
\author*[1,2]{\fnm{A. C.-L. } \sur{ Chian}}\email{abraham.chian@gmail.com}
\author[3]{\fnm{F. A.} \sur{ Borotto}}
\author[4,5]{\fnm{T.} \sur{ Hada}}
\author[6]{\fnm{R. A.} \sur{ Miranda}}
\author[7]{\fnm{P. R.} \sur{ Mu\~noz}}
\author[2,8]{\fnm{E. L.} \sur{ Rempel}}
\begin{itemize}
    {\tiny
\item[1] School of Mathematical Sciences,  University of Adelaide, Adelaide, SA 5005, Australia

\item[2]{  National Institute for Space Research (INPE), S\~ao Jos\'e dos Campos, SP 12227-010, Brazil

\item[3] Departamento de F\'isica,  Universidad de Concepci\'on, Concepci\'on,  Chile
\item[4] Department of Advanced Environmental Science \& Engineering,  Kyushu University}, Fukuoka, 8168580, Japan
\item[5] International Center for Space Weather Science \& Education (ICSWSE),  Kyushu University, Fukuoka, 8190382, Japan
\item[6] UnB-Gama Campus and Institute of Physics,  University of Bras\'ilia, Bras\'ilia, DF 70910-900, Brazil
\item[7] Departamento de F\'isica, Facultad de Ciencias,  Universidad de La Serena, Avenida Cisternas 1200, La Serena,  Chile
\item[8] Department of Mathematics,  Aeronautics Institute of Technology (ITA), S\~ao Jos\'e dos Campos, SP 12228-900, Brazil
    }
\end{itemize} 
{\small
* email abraham.chian@gmail.com
 }
\end{document}